\documentclass[12pt,hidelinks]{article}

\usepackage[T1]{fontenc}
\usepackage[utf8]{inputenc}
\usepackage{amsmath,amssymb,amsfonts,amsthm,mathtools}
\usepackage[separate-uncertainty=true,retain-unity-mantissa=false,load-configurations=abbreviations]{siunitx}[=v2]
\usepackage{slashed}
\usepackage{cancel}
\usepackage{graphicx}
\usepackage{booktabs,multicol,multirow}
\usepackage[shortlabels]{enumitem}
\usepackage{hyperref}
\usepackage[capitalize]{cleveref}
\usepackage{xspace}
\usepackage[noblocks]{authblk}
\usepackage[dvipsnames]{xcolor}
\usepackage{soul}
\usepackage[textsize=tiny]{todonotes}
\usepackage{subcaption}
\usepackage[normalem]{ulem}
\usepackage{nicefrac}
\usepackage{bm}
\usepackage{bbold}
\usepackage{dsfont}
\usepackage{makecell}

\definecolor{bg}{rgb}{0.95,0.95,0.95}

\usepackage[numbers,sort&compress]{natbib}
\usepackage{verbatim}

\newcommand{\cp}{\ensuremath{\mathcal{CP}}\xspace}

\newcommand{\sa}{{s_\alpha}}
\newcommand{\ca}{{c_\alpha}}

\newcommand{\aeff}{{\alpha_\text{eff}}}

\newcommand{\tev}{\,\, \mathrm{TeV}}
\newcommand{\gev}{\,\, \mathrm{GeV}}

\newcommand{\UFO}{\texttt{UFO}\xspace}

\newcommand{\madgraph}{{\texttt{MadGraph5\textunderscore aMC@NLO}}\xspace}
\newcommand{\pythia}{{\texttt{Pythia}\xspace}}
\newcommand{\higgstools}{{\texttt{HiggsTools}}\xspace}
\newcommand{\sushi}{{\texttt{SusHi}}\xspace}
\newcommand{\madanalysis}{{\texttt{MadAnalysis}}\xspace}
\newcommand{\delphes}{{\texttt{Delphes}}\xspace}

\newcommand{\chel}{\ensuremath{{c_{hel}}}}

\newcommand{\hOt}{\ensuremath{h_{1,\tree}}\xspace}
\newcommand{\hTt}{\ensuremath{h_{2,\tree}}\xspace}
\newcommand{\mhOt}{\ensuremath{m_{h_{1,\tree}}}\xspace}
\newcommand{\mhTt}{\ensuremath{m_{h_{2,\tree}}}\xspace}
\evensidemargin \oddsidemargin
\usepackage{scalerel}

\newcommand{\ttb}{\ensuremath{t\bar{t}}}

\newcommand{\tree}{\text{tree}}

\DeclarePairedDelimiter{\abs}{\lvert}{\rvert}

\newcommand{\ccite}[1]{Ref.\ \cite{#1}}
\newcommand{\ccites}[1]{Refs.\ \cite{#1}}

\begin{document}

\thispagestyle{empty}
\def\thefootnote{\fnsymbol{footnote}}

\begin{flushright}
\texttt{DESY-26-009}
\end{flushright}
\vspace{3em}
\begin{center}
{\large{\bf Complementarity of di-top and four-top searches\\[.3em]
in interpreting possible signals of new physics}}
\\
\vspace{3em}
 {
Henning Bahl$^{1}$
\footnotetext[0]{
bahl@thphys.uni-heidelberg.de,
philipp.gadow@uni-hamburg.de, 
romal.kumar@desy.de,
\hfill}\footnotetext[0]{
krisztian.peters@desy.de,
stylianou.panagiotis.1@ucy.ac.cy,
georg.weiglein@desy.de
\hfill},
Philipp Gadow$^{2}$,
Romal Kumar$^{3}$,
Krisztian Peters$^{3}$,\\[.3em]
Panagiotis Stylianou$^{4}$,
Georg Weiglein$^{3,5}$
 }\\[2em]
 {\sl $^1$ Institut für Theoretische Physik, Universität Heidelberg, Philosophenweg 16,\\ 61920 Heidelberg, Germany}\\[0.2em]
 {\sl $^{2}$ Institut f\"ur  Experimentalphysik, Universit\"at  Hamburg, Luruper Chaussee 149,\\ 22761 Hamburg, Germany}\\[0.2em]
 {\sl $^3$ Deutsches Elektronen-Synchrotron DESY, Notkestr.~85, 22607 Hamburg, Germany}\\[0.2em]
  {\sl $^{4}$ Department of Physics, University of Cyprus, P.O.\ Box 20537, 1678 Nicosia, Cyprus}\\[0.2em]
 {\sl $^{5}$ II.\  Institut f\"ur  Theoretische  Physik, Universit\"at  Hamburg, Luruper Chaussee 149,\\ 22761 Hamburg, Germany}\\
\def\thefootnote{\arabic{footnote}}
\setcounter{page}{0}
\setcounter{footnote}{0}
\end{center}
\vspace{2ex}

\begin{abstract}

Final states comprising two or more top quarks are important search channels at the Large Hadron Collider for scalar particles predicted in models of physics beyond the Standard Model. While the di-top final state profits from a higher signal cross section, it can be subject to intricate interference patterns. Besides the interference with the large QCD background, in case of the presence of more than one high-mass scalar also large signal--signal interference contributions can occur. We show that in such scenarios it is crucial to account for loop-level mixing for obtaining accurate exclusion bounds. We demonstrate how the interference patterns can obscure the interpretation of possible deviations from the Standard Model expectations. We show that the four-top final state, while giving rise to a smaller signal cross section, provides important complementary information due to its much smaller signal--background interference contributions. Thus, the results obtained from the four-top final state can be instrumental for pinpointing the underlying new physics scenario.

\end{abstract}
\setcounter{footnote}{0}
\renewcommand{\thefootnote}{\arabic{footnote}}

\newpage
\tableofcontents
\newpage

\section{Introduction}

The search for new fundamental scalar particles beyond the Standard Model (SM) is an important goal for the Large Hadron Collider (LHC). So far no clear evidence for such particles has been found. In the next years, the amount of data will, however, increase very significantly. Moreover, advanced analysis techniques will allow further improvements of the sensitivity. This is in particular true for complex high-dimensional final states.

With these prospects in mind, the search for Beyond-the-Standard-Model (BSM) scalars in final states comprising several top quarks becomes increasingly promising. These final states are also theoretically well-motivated since in many BSM scenarios the largest coupling of BSM scalars is their coupling to the top quark, as for the SM Higgs boson.

The di-top production process is of particular interest because of its large cross section at the LHC. This process is dominated by the gluon-induced production, for which also non-perturbative effects were found to play an important role near the di-top production threshold~\cite{CMS:2025kzt,ATLAS:2026dbe,Fuks:2021xje}. It is well known that due to the interference of the signal --- i.e., the resonant production of one or more BSM scalar(s) --- with the QCD background, BSM scalars do not manifest themselves in the form of a single peak in the invariant mass spectrum but in the form of a characteristic peak--dip structure~\cite{Gaemers:1984sj,Dicus:1994bm,Frederix:2007gi,Craig:2015jba,Jung:2015gta,Bernreuther:2015fts,Carena:2016npr,Hespel:2016qaf,BuarqueFranzosi:2017jrj,Djouadi:2019cbm}. The existing experimental searches specifically focus on this peak--dip signature~\cite{ATLAS:2017snw,CMS:2019pzc,CMS-PAS-HIG-22-013,ATLAS:2024vxm,CMS:2025kzt,ATLAS:2025mvr}. 

Many BSM models, however, contain more than one BSM scalar as for example the widely studied Two-Higgs Doublet Model (2HDM) --- see~\ccites{Gunion:1989we,Branco:2011iw,Ivanov:2017dad} for reviews. The possibilities for several scalars and \cp-violating effects also exist in the Higgs sector of the Minimal Supersymmetric extension of the SM (MSSM), for which LHC di-top search benchmarks have been defined~\cite{Bagnaschi:2018ofa,Bahl:2019ago}. The 2HDM predicts two neutral BSM scalars that predominantly couple to top quarks if the vacuum expectation values of the two doublets are close to each other. The complex version of the 2HDM (c2HDM) can, moreover, accommodate \cp violation, which leads to mixing between the two BSM scalars. In such a scenario, the presence of interference between the two scalars as well as between the background and each of the two scalars can significantly modify the di-top invariant mass spectra compared to the peak--dip structure expected for a single scalar~\cite{Bernreuther:2015fts,Carena:2016npr,Bahl:2025you}. In case of a deviation from the SM expectation in the di-top channel, the possible interference patterns will make it difficult to relate the observed signature to a specific underlying new physics scenario. As shown in~\ccite{Bahl:2025you}, extending the formalism worked out in~\ccites{Fuchs:2014ola,Fuchs:2016swt,Fuchs:2017wkq}, it is crucial to incorporate loop-level mixing contributions for describing scenarios with two or more mixing BSM scalars. These contributions can have an important impact on the invariant mass spectrum.

The presence of BSM scalars with large couplings to top quarks manifests itself, however, not only in the di-top final state. In the four-top final state, the signal--signal interference contributions are also present, while the relative importance of the signal--background interference contributions is much reduced compared to the di-top final state. Four-top production has a much smaller cross-section than di-top production and has only recently been observed~\cite{ATLAS:2023ajo,CMS:2023ftu}. The relative size of BSM contributions can, however, be larger than for di-top production motivating respective experimental searches~\cite{CMS-TOP-18-003,ATLAS:2022rws,ATLAS:2024bsm,ATLAS:2024jja}. The fact that the interference patterns differ significantly from the di-top final state due to the much smaller SM contribution and the overall different kinematics add further motivation for studying the four-top final state. Specifically, for the case of a single BSM scalar in the four-top final state it is known that the interference with the SM background is negligible, resulting in a clean peak signature~\cite{Anisha:2023xmh,Craig:2017}. This complementarity with the di-top final state could make the four-top final state an ideal channel for determining the physical origin of an observed deviation in the di-top final state.

In this paper, we investigate the interplay between the di-top and four-top channels in detail. Moreover, we quantify the effect of loop-level mixing on the experimental sensitivity for both the di-top and four-top channels. The paper is structured as follows. In~\cref{sec:toy-model-zfac}, we discuss loop-level mixing and investigate the size of the effects in a toy model. In~\cref{sec:setup}, we describe the reinterpretation of the expected sensitivities of experimental di-top and four-top searches that we use to study the interplay between di-top and four-top channels at the LHC, as well as our Monte Carlo (MC) event generation setup. We discuss the interference patterns for both processes and the impact of loop-level mixing in~\cref{sec:top_interference}. In~\cref{sec:tt_tttt_interplay}, we investigate the interplay of both processes. We present conclusions in~\cref{sec:conclusions} and provide additional numerical results for the loop-level mixing in the toy model in~\cref{app:eff_mix} and for the significance scans in~\cref{app:sig_scans}.

\section{Loop-level mixing and \texorpdfstring{$Z$}{Z}-factor formalism}
\label{sec:toy-model-zfac}

For our analysis, we work in a simplified model framework, which extends the SM by two scalars. Their interaction with top quarks is given in the tree-level mass eigenstate basis by
\begin{align}
    \label{eq:Lyuk}
	\mathcal{L}_\text{yuk} = -\sum_{j=1}^{2}\dfrac{y_t^\text{SM}}{\sqrt{2}}\,\bar{t}\left(c_{t,j} + i\gamma_5 \tilde{c}_{t,j}\right)t\, h_{j,\tree} \;.
\end{align}
Here, $y_t^\text{SM}$ is the SM top-Yukawa coupling. The 
coefficients
$c_{t,j}$ are the \cp-even coupling modifiers and $\tilde c_{t,j}$ the \cp-odd ones.

Loop-level mixing between two tree-level mass eigenstates $h_{1,\tree}$ and $h_{2,\tree}$ is induced by self-energy corrections. These contributions in particular give rise to the fact that the loop-corrected masses differ from the tree-level ones. Since the poles of the propagators occur at the physical, i.e.\ loop-corrected, masses, the incorporation of these contributions is crucial for a reliable theoretical description. We can write amplitudes with an intermediary scalar line in the form 
\begin{align}
    \label{eq:prod_decay_delta_decomposition}
    \mathcal{A} = \Gamma_{i,\text{prod}} \Delta_{ij}(p^2) \Gamma_{j,\text{decay}}\;,
\end{align}
where $\Gamma_{i,\text{prod}}$ is the production amplitude of the scalars and $\Gamma_{i,\text{decay}}$ is their decay part. The propagator matrix
\begin{align}
    \label{expr-propagator-matrix}
    \bm{\Delta}_{ij} (p^2) = \begin{pmatrix}
        \Delta_{h_{1,\tree}h_{1,\tree}} & \Delta_{h_{1,\tree}h_{2,\tree}} \\
        \Delta_{h_{2,\tree}h_{1,\tree}} & \Delta_{h_{2,\tree}h_{2,\tree}}
        \end{pmatrix}_{ij}
        := - \left(\hat{\bm{\Gamma}}_{ij}(p^2)\right)^{-1} \,
\end{align}
is given by the inverse of the renormalised, one-particle irreducible (1PI) two-point vertex function $\hat{\bm{\Gamma}}_{ij}(p^2)$, which we can write in the form
\begin{align}
    \hat{\bm{\Gamma}}_{ij}(p^2) = i\left[(p^2 - m_i^2)\delta_{ij} + \hat{\Sigma}_{ij}(p^2)\right] \,.
    \label{eq:vertexfct}
\end{align}
Here, $m_i$ are the tree-level masses and $\hat{\Sigma}_{ij}(p^2)$ are the renormalised self-energies of the two scalars that are expressed in terms of their tree-level mass eigenstates. 

While it is possible in principle to apply~\cref{eq:prod_decay_delta_decomposition} directly, it is useful to express $\Gamma_{i,\text{prod}}$, $\Gamma_{i,\text{decay}}$ and the propagator contribution in terms of the (loop-corrected) mass eigenstates instead of the tree-level states. This can be done via the so-called $Z$-factor formalism which we will briefly summarise next. We furthermore introduce the effective mixing angle approximation, which is sometimes used in this context.

\subsection{\texorpdfstring{$Z$}{Z}-factor formalism}
\label{sec:z-factors}

The $Z$-factor formalism has been developed in detail for the treatment of unstable particles in~\ccites{Fuchs:2017wkq,Fuchs:2016swt} (see~\ccites{Dabelstein:1995js,Pilaftsis:1997dr,Heinemeyer:2000fa,Heinemeyer:2001iy,Ellis:2004fs} for earlier works). Following the discussion in~\ccite{Fuchs:2016swt}, the elements of the diagonal propagators $\Delta_{ii}(p^2)$ and off-diagonal propagators $\Delta_{ij}(p^2)$ (with $i \neq j$) can be written as
\begin{align}
    \label{expr:propagator-matrix-element}
    \Delta_{ii}(p^2) & = \dfrac{i\left[D_j(p^2) + \hat{\Sigma}_{jj}(p^2) \right]}{\left[D_i(p^2) + \hat{\Sigma}_{ii}(p^2) \right] \left[D_j(p^2) + \hat{\Sigma}_{jj}(p^2) \right] - \hat{\Sigma}_{ij}^2(p^2)} = \dfrac{i}{p^2 - m_i^2 + \hat{\Sigma}_{ii}^{\text{eff}}(p^2)} \,,\\
    \Delta_{ij}(p^2) & = \dfrac{-i\hat{\Sigma}_{ij}(p^2)}{\left[D_i(p^2) + \hat{\Sigma}_{ii}(p^2) \right] \left[D_j(p^2) + \hat{\Sigma}_{jj}(p^2) \right] - \hat{\Sigma}_{ij}^2(p^2)} \qquad (i\neq j) \,,
\end{align}
where
\begin{align}
    \hat{\Sigma}_{ii}^{\text{eff}}(p^2) = \hat{\Sigma}_{ii}(p^2) - \dfrac{\hat{\Sigma}_{ij}^2(p^2)}{D_j(p^2) + \hat{\Sigma}_{jj}(p^2)} \,, \qquad \text{and} \qquad D_i(p^2) = p^2 - m_i^2 \;.
\end{align}
The diagonal ($\Delta_{ii}$) and the off-diagonal ($\Delta_{ij}$) components of the propagator matrix given in~\cref{expr-propagator-matrix} each have two complex poles, denoted as $\mathcal{M}_{h_1}^2$ and $\mathcal{M}_{h_2}^2$. In contrast to the effective mixing angle approach described below, the momentum dependence is fully taken into account. We order the complex poles according to their real part --- i.e., $\text{Re}\,(\mathcal{M}_{h_1}^2) \leq \text{Re}\,(\mathcal{M}_{h_2}^2)$.

Moreover, the propagator matrix is used to obtain wave-function normalisation factors which ensure the proper normalisation of the S~matrix for external particles, according to the application of the well-known Lehmann-Symanzik-Zimmermann (LSZ) reduction formula~\cite{Lehmann:1954rq} to the case of unstable external particles. Associating the indices $(a, b)$ with the loop-corrected mass eigenstates $(h_1, h_2)$, we can write the wave-function normalisation factor for $i$--$j$ mixing on the scalar line evaluated at the pole $\mathcal{M}_a^2$ as a product of the overall normalisation factor $(\hat{Z}^{a}_i)^{1/2}$ --- corresponding to the LSZ factor in the case of no mixing --- times the on-shell transition ratio $\hat{Z}^{a}_{ij}$, where
\begin{align}
    \label{expr:os-transition-zfac}
    \hat{Z}^a_i = \left. \left(1 + \dfrac{\partial \hat \Sigma^{\text{eff}}_{ii}(p^2)}{\partial p^2}\right)^{-1}\right|_{p^2 = \mathcal{M}_a^2} \,, 
    \qquad \text{and} \qquad
    \hat{Z}^{a}_{ij} = \left. \dfrac{\Delta_{ij}(p^2)}{\Delta_{ii}(p^2)}\right|_{p^2 = \mathcal{M}_a^2} \,.
\end{align}
It has been shown that different assignments between the tree-level mass eigenstates and the loop-corrected mass eigenstates are physically equivalent~\cite{Fuchs:2016swt}. For the present case of only two mixing scalars, it is convenient to associate the lighter and heavier of the tree-level mass eigenstates ($h_{1,\tree}, h_{2,\tree}$) with the lighter and heavier one of the loop-corrected mass eigenstates ($h_1, h_2$), respectively. This assignment denoted as $((h_{1,\tree}, h_1), (h_{2,\tree}, h_2))$ implies the evaluation of
\begin{align}
    \label{expr:zfac_assignment_1}
    \hat{Z}^{h_1}_{h_{1,\tree}} \,, \quad
    \hat{Z}^{h_1}_{h_{1,\tree} h_{2,\tree}}
    \quad\text{at}\quad p^2 = \mathcal{M}_{h_1}^2 
\end{align}
and
\begin{align}
    \label{expr:zfac_assignment_2}
    \hat{Z}^{h_2}_{h_{2,\tree}} \,, \quad
    \hat{Z}^{h_2}_{h_{2,\tree} h_{1,\tree}}
    \quad\text{at}\quad p^2 = \mathcal{M}_{h_2}^2 \,.
\end{align}
In this chosen assignment and notation, $\hat{Z}^{h_1}_{h_{1,\tree}h_{1,\tree}} = \hat{Z}^{h_2}_{h_{2,\tree}h_{2,\tree}} =~1 \,.$

The normalisation factors $(\hat{Z}^a_i)^{1/2}$ and transition ratios $\hat{Z}^{a}_{ij}$ described in~\cref{expr:zfac_assignment_1,expr:zfac_assignment_2} can be arranged into a so-called $\bm{{Z}}$-matrix which is a non-unitary complex matrix.\footnote{This is related to imaginary parts appearing in the propagators of unstable particles.} The specific $\bm{{Z}}$-matrix for the chosen assignment between the tree-level and loop-corrected mass eigenstates can be written as
\begin{align}
    \label{expr-z-matrix}
    \bm{{Z}} = \begin{pmatrix}
        \sqrt{\hat{Z}^{h_1}_{h_{1,\tree}}}\hat{Z}^{h_1}_{h_{1,\tree}h_{1,\tree}} & \sqrt{\hat{Z}^{h_1}_{h_{1,\tree}}}\hat{Z}^{h_1}_{h_{1,\tree}h_{2,\tree}} \\
        \sqrt{\hat{Z}^{h_2}_{h_{2,\tree}}}\hat{Z}^{h_2}_{h_{2,\tree}h_{1,\tree}} & \sqrt{\hat{Z}^{h_2}_{h_{2,\tree}}}\hat{Z}^{h_2}_{h_{2,\tree}h_{2,\tree}}
    \end{pmatrix} \equiv
    \begin{pmatrix}
        \bm{{Z}}^{h_1}_{h_{1,\tree}} & \bm{{Z}}^{h_1}_{h_{2,\tree}} \\
        \bm{{Z}}^{h_2}_{h_{1,\tree}} & \bm{{Z}}^{h_2}_{h_{2,\tree}} \\
    \end{pmatrix} \,.
\end{align}

As shown in~\ccite{Fuchs:2016swt}, the loop-corrected internal propagator of~\cref{expr:propagator-matrix-element} taking into account the relevant higher-order contributions is well approximated using Breit-Wigner propagators for the (loop-corrected) mass eigenstates and the $\bm{{Z}}$-matrix,
\begin{align}
    \label{expr-bw-z-factor}
    \Delta_{ii}(p^2) \simeq \sum_{a=h_1,h_2}^{} (\bm{{Z}}^{a}_{i})^2\Delta^{\text{BW}}_a(p^2) \,, \qquad\text{and}\qquad
    \Delta_{ij}(p^2) \simeq \sum_{a=h_1,h_2}^{} \bm{{Z}}^{a}_{i}\Delta^{\text{BW}}_a(p^2)\bm{{Z}}^{a}_{j} \,.
\end{align}
Here, $\Delta^{\text{BW}}_a(p^2)$ is the Breit-Wigner propagator
\begin{align}
    \Delta^{\text{BW}}_a(p^2) = \dfrac{i}{p^2 - \mathcal{M}_a^2} = \dfrac{i}{p^2 - M^2_{a} + i M_a \Gamma_a} \,,
\end{align}
with the loop-corrected mass $M_a$ and the total width $\Gamma_a$.

As detailed in~\ccite{Bahl:2025you}, this can be used to incorporate the effects of loop-level mixing via the complex poles in conjunction with replacing the Yukawa-modifiers by
\begin{subequations}
    \label{expr-ct-ctt-replacements}
    \begin{align}
    \label{expr-ct-ctt-replacements-ct1} c_{t,1} & \rightarrow \bm{Z}^{h_1}_{h_{1,\tree}}c_{t,1} + \bm{Z}^{h_1}_{h_{2,\tree}}c_{t,2} \,, \\
    \label{expr-ct-ctt-replacements-ctt1} \tilde{c}_{t,1} & \rightarrow \bm{Z}^{h_1}_{h_{1,\tree}}\tilde{c}_{t,1} + \bm{Z}^{h_1}_{h_{2,\tree}}\tilde{c}_{t,2} \,, \\
    \label{expr-ct-ctt-replacements-ct2} c_{t,2} & \rightarrow \bm{Z}^{h_2}_{h_{2,\tree}}c_{t,2} + \bm{Z}^{h_2}_{h_{1,\tree}}c_{t,1} \,, \\
    \label{expr-ct-ctt-replacements-ctt2} \tilde{c}_{t,2} & \rightarrow \bm{Z}^{h_2}_{h_{2,\tree}}\tilde{c}_{t,2} + \bm{Z}^{h_2}_{h_{1,\tree}}\tilde{c}_{t,1} \,.
    \end{align}
\end{subequations}
Notably, the $\bm{Z}$-matrix elements are complex numbers which can lead to additional phases in the scattering amplitudes beyond the ones arising directly from loop integrals.

\subsection{Effective mixing angle approximation}
\label{sec:effective_mixing}

In the effective-mixing-angle approximation, the momentum at which the self-energies in \cref{eq:vertexfct} are evaluated is fixed~\cite{Heinemeyer:2000fa}. While $p^2_\text{fixed} = 0$ is the simplest choice that was used in~\ccite{Heinemeyer:2000fa}, in principle also other choices are possible.

After fixing the momentum for which the loop corrections are evaluated, the loop corrections give rise to a simple basis rotation of the tree-level mass eigenstates. In particular, the loop-corrected poles in this approximation are given by the solution of
\begin{align}
    \text{det}
    \begin{pmatrix}
        p^2 - \mhOt^2 + \hat\Sigma_{\hOt\hOt}(p_\text{fixed}^2) & \hat\Sigma_{\hOt\hTt}(p_\text{fixed}^2) \\
        \hat\Sigma_{\hOt\hTt}(p_\text{fixed}^2) & p^2 - \mhTt^2 + \hat\Sigma_{\hTt\hTt}(p_\text{fixed}^2) \\
    \end{pmatrix}
    = 0 \;.
\end{align}
Moreover, we can define the effective mixing angle $\Delta\alpha$ as the angle which diagonalises this matrix,
\begin{align}
    \tan\Delta\alpha = \frac{2\hat\Sigma_{\hOt\hTt}(p_\text{fixed}^2)}{\Delta M^2 + \sqrt{(\Delta M^2)^2 + 4 \hat\Sigma_{\hOt\hTt}^2(p_\text{fixed}^2)}} \;,
\end{align}
with
\begin{align}
    \Delta M^2 = (\mhTt^2 - \hat\Sigma_{\hTt\hTt}(p_\text{fixed}^2)) - (\mhOt^2 - \hat\Sigma_{\hOt\hOt}(p_\text{fixed}^2)) \;.
\end{align}
This angle relates the tree-level and loop-level states in the zero-momentum approximation:
\begin{align}
    \begin{pmatrix}
        h_1 \\ h_2
    \end{pmatrix}
    = 
    \begin{pmatrix}
        - s_{\Delta\alpha} & c_{\Delta\alpha} \\
          c_{\Delta\alpha} & s_{\Delta\alpha} 
    \end{pmatrix}
    \begin{pmatrix}
        \hOt \\ \hTt
    \end{pmatrix}\;,
\end{align}
where we introduced the short-hand notation $s_\gamma\equiv\sin\gamma$ and $c_\gamma\equiv\cos\gamma$.

\subsection{Comparison for a toy model}
\label{sec:toy_model}

For comparing the different approaches, we consider a toy model with two scalars \hOt and \hTt, which couple to top quarks via
\begin{align}
    \mathcal{L}_\text{top-Yuk} = \frac{y_t}{\sqrt{2}}\Big(\ca \hOt \bar t t + \sa \hTt \bar t t\Big)\;.
\end{align}
Here, $\alpha$ is a tree-level mixing angle controlling the strength of the Yukawa couplings of \hOt and \hTt. $y_t$ is the SM top-Yukawa coupling.

This setup allows us to factor out the mixing angles from the renormalised self-energies,
\begin{subequations}
\begin{align}
    \hat\Sigma_{\hOt\hOt}(p^2) &= c_\alpha^2 \hat\Sigma(p^2)\;, \\
    \hat\Sigma_{\hOt\hTt}(p^2) &= \sa\ca \hat\Sigma(p^2)\;, \\
    \hat\Sigma_{\hTt\hTt}(p^2) &= s_\alpha^2 \hat\Sigma(p^2)\;.
\end{align}
\end{subequations}
For our numerical comparison we consider the case $\alpha = \pi/4$ for which $s_\alpha = c_\alpha = 1/\sqrt{2}$.

As an exemplary toy process, we consider $t\bar t \to h_{1,2} \to t\bar t$, which we here consider as a simplified proxy for the $gg \to h_{1,2} \to t\bar{t}$ and $gg \to t\bar t h_{1,2} \to t\bar t t\bar t $ processes considered later. We are in particular interested in the nearly mass-degenerate case for which $\mhOt \rightarrow \mhTt$, since in this limit interference effects could become large. As discussed in~\ccite{Sakurai:2022cki,LoChiatto:2024guj}, in this limit the so-called ``quantum Zeno effect'' occurs in which the width of one of the scalars that mix with each other goes to zero. This quantum Zeno effect has also been discussed for $t\bar t$ production in~\ccite{Bahl:2025you}.

Here, we go beyond~\ccite{Bahl:2025you} demonstrating analytically and numerically how the zero-width resonance appearing in the limit $\mhTt\to \mhOt$ is regulated. Following up on this discussion, we will show in the next sections how this directly affects experimental bounds on BSM scalars from searches for multi-top final states.

\subsubsection*{Effective mixing angle approach}

We consider here first the effective mixing-angle approach involving the approximation $\hat\Sigma(p^2)\simeq \hat\Sigma(p^2_\text{fixed})$. We choose the mean of the tree-level masses for the fixed momentum, $p^2_\text{fixed} = (\mhOt + \mhTt)^2/4$. Then the complex poles can easily be computed as
\begin{align}
    \mathcal{M}_{h_{1,2}}^2 = \frac{1}{2}\bigg(&\mhOt^2 + \mhTt^2 - \hat\Sigma(p^2_\text{fixed}) \nonumber\\
    &\mp \sqrt{(\mhOt^2 - \mhTt^2)^2 + \hat\Sigma^2(p^2_\text{fixed}) - 2(\mhOt^2 - \mhTt^2) c_{2\alpha} \hat\Sigma(p^2_\text{fixed})}\bigg) \;.
\end{align}
In the limit $\Delta m^2 = \mhTt^2 - \mhOt^2 \sim 0$, the poles and loop-level shift of the mixing angle become\footnote{Here, we assume $\text{Re}\,\hat\Sigma(p^2_\text{fixed})>0$. If $\text{Re}\,\hat\Sigma(p^2_\text{fixed})<0$ the roles of $h_1$ and $h_2$ are interchanged.}
\begin{align}
    \mathcal{M}_{h_1}^2 & \equiv M_{h_1}^2 - i M_{h_1}\Gamma_{h_1} = \mhOt^2  + c_\alpha^2 \Delta m^2 + \mathcal{O}(\Delta m^4)\;,\\
    \mathcal{M}_{h_2}^2 & \equiv M_{h_2}^2 - i M_{h_2}\Gamma_{h_2} = \mhTt^2 - \hat\Sigma(p^2_\text{fixed}) + s_\alpha^2 \Delta m^2 + \mathcal{O}(\Delta m^4)\;,\\
    \tan\Delta\alpha &= \tan\alpha \left(1 - \frac{\Delta m^2}{\hat\Sigma(p^2_\text{fixed})} + \mathcal{O}(\Delta m^4)\right)\;.
\end{align}
This yields for the effective mixing angle 
\begin{align}
    \aeff = \alpha - \Delta\alpha = \frac{s_{2\alpha}}{2\hat\Sigma(p^2_\text{fixed})}\Delta m^2 + \mathcal{O}(\Delta m^4)\;.
\end{align}
The corresponding couplings of the loop-level states are given by
\begin{align}
    c(h_1 t\bar t) &= y_t(- s_{\Delta\alpha} \ca + c_{\Delta\alpha} s_a) = y_t\sin(\alpha - \Delta\alpha) = y_t\sin(\aeff)\;,\\
    c(h_2 t\bar t) &= y_t(c_{\Delta\alpha} \ca + s_{\Delta\alpha} s_a) = y_t\cos(\alpha - \Delta\alpha) = y_t\cos(\aeff) \;.
\end{align}
Consequently, in the limit $\mhTt^2 \to \mhOt^2$, only $h_2$ couples to top quarks.

If we now consider the process $t\bar t \to h_{1,2} \to t\bar t$, the squared amplitude is proportional to
\begin{align}
    \label{eq:M2_tt_to_tt}
    |\mathcal{M}|^2 \propto
    \left| \frac{s_\aeff^2}{s - M_{h_1}^2 + i M_{h_1}\Gamma_{h_1}} + \frac{c_\aeff^2}{s - M_{h_2}^2 + i M_{h_2}\Gamma_{h_2}}\right|^2 \;.
\end{align}
Integrating one of the Breit-Wigner propagators over $s$ yields
\begin{align}
    \int_0^\infty ds \left| \frac{1}{s - M^2 + i M\Gamma}\right|^2 = \frac{\pi + 2\arctan\frac{M}{\Gamma}}{2 M \Gamma} \simeq \frac{\pi}{M \Gamma} - \frac{1}{M^2} + \mathcal{O}\left(\frac{\Gamma}{M^3}\right)\;.
\end{align}
This implies that if the width goes to zero, the integrated Breit-Wigner propagator diverges. As we see in~\cref{eq:M2_tt_to_tt}, this divergence is regulated at the level of the squared matrix element since $s_\aeff^2\to 0$ as $\Gamma_{h_1}\to 0$ in the limit $\mhOt \to \mhTt$.

\subsubsection*{\texorpdfstring{$Z$}{Z}-factor formalism}

For $\alpha = \pi/4$ and $\mhOt = \mhTt$, the effective $\hOt\hOt$ self-energy simplifies to
\begin{align}
    \hat\Sigma^\text{eff}_{\hOt\hOt}(p^2) = (p^2 - \mhOt^2)\cdot\frac{c_\alpha^2\hat\Sigma(p^2)}{p^2 - \mhOt^2 + s_\alpha^2\hat\Sigma(p^2)} \;.
\end{align}
It follows from~\cref{expr:propagator-matrix-element} that $\mhOt^2$ is one of the physical poles. For this pole, we get
\begin{align}
    \frac{\partial\hat\Sigma^\text{eff}_{\hOt\hOt}}{\partial p^2}\bigg|_{p^2 = \mhOt^2} &= \frac{c_\alpha^2}{s_\alpha^2}
\end{align}
and 
\begin{align}
    \sqrt{\hat Z_{\hOt}^{h_1}} &= s_\alpha, 
    \hspace{1cm}
    \hat Z_{{h_{1,\tree}h_{2,\tree}}}^{h_1}  =  - \frac{c_\alpha}{s_\alpha} \;.
\end{align}
Consequently, the effective loop-level top-Yukawa coupling of $h_1$, given by
\begin{align}
    c(h_1 t\bar t)=  \bm{Z}^{h_1}_{h_{1,\tree}} c_\alpha + \bm{Z}^{h_1}_{h_{2,\tree}} s_\alpha \;,
\end{align}
vanishes in the limit $\mhTt\to \mhOt$ preventing the integrated $t\bar t \to h_{1,2} \to t\bar t$ cross-section to grow to infinity as $\Gamma_{\hOt}\to 0$. This regularisation mechanism is the same as in the effective mixing angle approach. We also checked that the same happens for the case where the full propagator matrix is used.

\subsubsection*{Numerical comparison}

Next, we numerically compare the three different approaches. For this study, we fix $\alpha = \pi/4$, corresponding to a maximally mixed situation at the tree level, and $m_{h_{2,\tree}} = 500\gev$.
\begin{figure}
    \centering
    \includegraphics[width=0.49\linewidth]{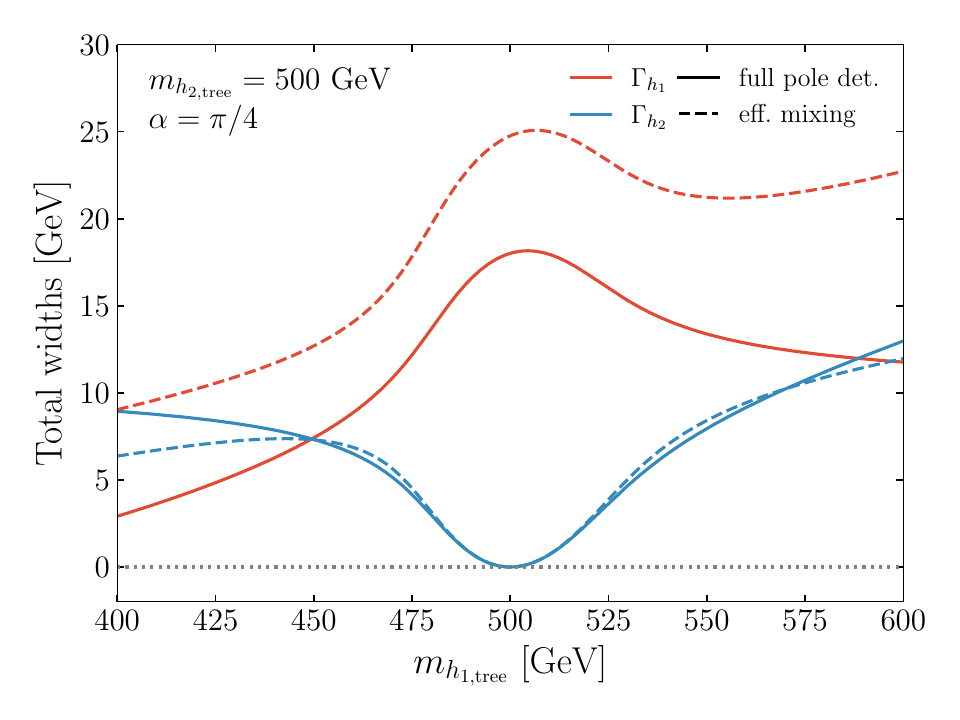}
    \includegraphics[width=0.49\linewidth]{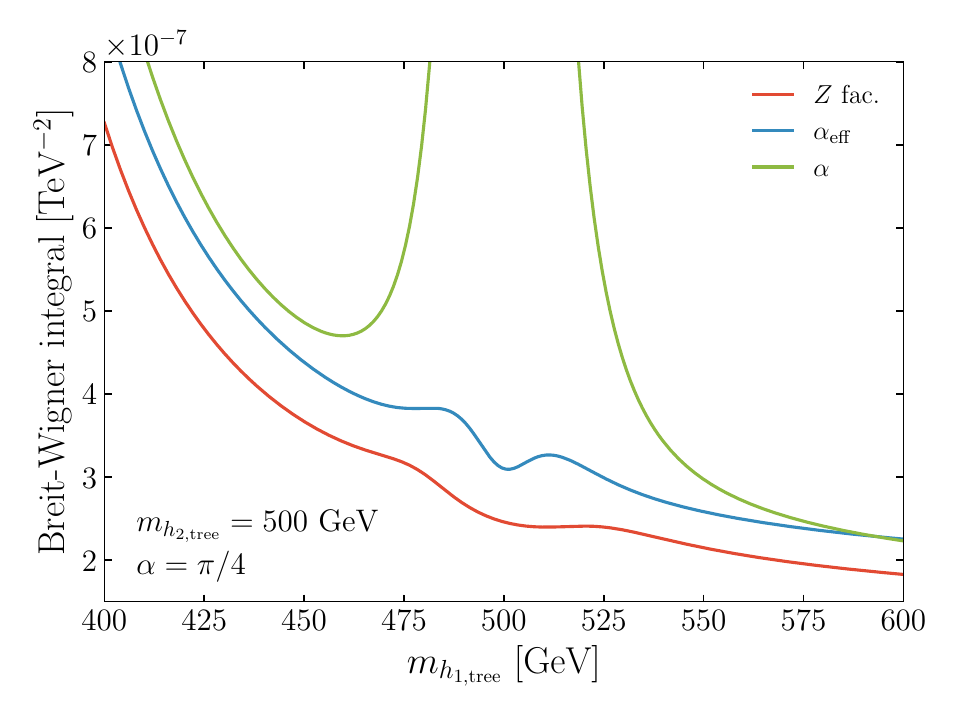}
    \caption{Left: Total width of $h_{1,2}$ as a function of $m_{h_{1,\tree}}$ comparing the results of the full pole determination to the effective mixing angle approach. Right: Integral over the squared sum of the Breit-Wigner propagators as a function of $m_{h_{1,\tree}}$ using either the tree-level mixing angle, the effective mixing angle, or the $Z$-factor formalism for the couplings of the mass eigenstates.}
    \label{fig:toy_widths_BWints}
\end{figure}

In the left panel of~\cref{fig:toy_widths_BWints}, we display the total widths of the mass eigenstates as a function of $m_{h_{1,\tree}}$. While the effective-mixing-angle approach approximates the width of the heavier mass eigenstate relatively well, the width of the lighter mass eigenstates is overestimated by a factor of $\sim 2$. This is a consequence of evaluating the self-energies at the fixed momentum $p^2 = \frac{1}{4}(m_{h_{1,\tree}} + m_{h_{2,\tree}})^2$. We provide additional figures showing the loop-corrected masses and the effective mixing angle as a function of $m_{h_{1,\tree}}$ in~\cref{app:eff_mix}.

Here, we focus on the integral over the squared sum of the Breit-Wigner propagators serving as a proxy for the total cross-section. In the right panel of~\cref{fig:toy_widths_BWints}, we show this integral as a function of $m_{h_{1,\tree}}$ using either the tree-level mixing angle, the effective mixing angle, or the $Z$-factor formalism for the couplings of the mass eigenstates. Explicitly, the integrals are defined as
\begin{alignat}{2}
    &\text{Breit-Wigner integral}\bigg|_\alpha &&= \int_0^{1\tev^2} ds \left|\frac{s_\alpha^2}{s - M_{h_1}^2 + i M_{h_1}\Gamma_{h_1}} + \frac{c_\alpha^2}{s - M_{h_2}^2 + i M_{h_2}\Gamma_{h_2}}\right|^2 \;,\\
    &\text{Breit-Wigner integral}\bigg|_{\aeff} &&= \int_0^{1\tev^2} ds \left|\frac{s_\aeff^2}{s - M_{h_1}^2 + i M_{h_1}\Gamma_{h_1}} + \frac{c_\aeff^2}{s - M_{h_2}^2 + i M_{h_2}\Gamma_{h_2}}\right|^2 \;,\\
    &\text{Breit-Wigner integral}\bigg|_{Z\text{-factor}} &&= \int_0^{1\tev^2} ds \left|\frac{(\bm{Z}^{h_1}_{h_{1,\tree}} s_\alpha + \bm{Z}^{h_1}_{h_{2,\tree}} c_\alpha)^2}{s - M_{h_1}^2 + i M_{h_1}\Gamma_{h_1}} \right. \nonumber\\
    &&&\left.\hspace{2.6cm} + \frac{(\bm{Z}^{h_2}_{h_{2,\tree}} c_\alpha + \bm{Z}^{h_2}_{h_{1,\tree}} s_\alpha)^2}{s - M_{h_2}^2 + i M_{h_2}\Gamma_{h_2}}\right|^2 \;.
\end{alignat}
The shown results do only mildly depend on the chosen upper cut-off of 1~TeV. 

Looking at the results in the right panel of~\cref{fig:toy_widths_BWints}, we see that close to the crossing point, at which $\mhOt \simeq \mhTt$, using the tree-level mixing angle leads to a significant overestimate of the integral. The sharp increase of the integral close to the crossing point is prevented in the effective mixing angle formalism, which, however, still slightly overestimates the integral as a consequence of the deviation in the prediction for the decay width $\Gamma_{h_1}$ as shown in the left panel of~\cref{fig:toy_widths_BWints}. The $Z$-factor curve is the smoothest result with only very weak interference patterns being visible close to the crossing point.

In conclusion, the proper treatment of loop-level mixing is essential to obtain physically meaningful results. The $Z$-factor formalism takes into account the momentum dependence and provides a consistent treatment of the complex poles that are associated with unstable particles. As shown in~\ccite{Fuchs:2016swt}, expressing the propagators in terms of the (loop-corrected) mass eigenstates via the $Z$-factor formalism yields a very good approximation of the full propagator matrix. It furthermore enables a straightforward implementation into Monte-Carlo event generators. The effective mixing angle approach provides a simple approximation of the loop-level mixing contributions.

\section{Event generation setup and reinterpretation of the expected sensitivities of the employed LHC searches}
\label{sec:setup}

After discussing the effect of loop-level mixing for a toy model, we now turn to top resonance searches
in di-top and four-top final states. Searches for scalar resonances in di-top and four-top channels are performed at the LHC~\cite{Evans:2008} by both ATLAS~\cite{ATLAS:2017snw,ATLAS:2024vxm} and CMS~\cite{CMS:2019pzc,CMS:2025dzq}.\footnote{See also~\ccite{ATLAS:2024itc} for a comprehensive review of ATLAS searches for additional scalar particles and exotic Higgs-boson decays based on the full LHC Run-2 dataset.} We proceed by describing the characteristic features of the di-top and four-top channels, detailing the event generation setup and describing searches performed by the CMS collaboration that we will use to re-cast the expected sensitivities such that they can be applied to the considered benchmark models specified below. Our analysis will make use of the $Z$-factor formalism described in~\cref{sec:z-factors}.

We implement an analysis for each channel motivated by the CMS searches by taking the background prediction of the results from the experimental searches, while we perform a simulation of events in order to obtain the signal contributions. We use \UFO model files with up to two additional scalars and the $\bm{Z}$-matrix implemented as described in~\cref{sec:z-factors}. Details on the implementation can be found in~\ccite{Bahl:2025you}.

The significance of the respective channels is evaluated following the approach for estimating the discovery sensitivity for a counting experiment with background uncertainty~\cite{Cowan:2012medsig}.

\subsection{Searches for new scalar particles in di-top final states}
\label{sec:tt}

This channel is known to be impacted by sizeable interference effects between the large SM background and new physics which can lead to peak--dip or dip--peak structures instead of Breit-Wigner resonances in the invariant mass distribution, $m_{t \bar{t}}$. The actual shape of the distribution provides important information about a possible signal. While the results of the experimental searches are commonly presented
in terms of a single \cp-even or \cp-odd neutral scalar, they can be reinterpreted for the case of two \cp-mixed neutral states decaying to $t \bar{t}$. Because of the possible impact of such scenarios on the invariant mass distribution, experimental results in the di-top final state may be difficult to associate with a specific signal of new physics.

We proceed by implementing an analysis motivated from the CMS search~\cite{CMS:2019pzc} using \SI{13}{\tera\electronvolt} \(pp\) collision data corresponding to an integrated luminosity of \SI{35.9}{\per\femto\barn}, focusing on the $p p \rightarrow h_{1,2} \rightarrow t \bar{t}$ process with fully-leptonic decays of the top-quarks, $t \rightarrow b \ell^+ \nu_{\ell}$, and similarly for $\bar{t}$. The fully-leptonic final state allows one to fully exploit observables constructed from spin-correlations of the $t \bar{t}$ system, providing an improved separation between the QCD background and the signal comprising a neutral scalar in the $s$-channel, leading to an increased statistical significance. 

The extraction of the relevant observables proceeds by first defining an appropriate coordinate system by the unit vectors $\hat{k}$, $\hat{r}$, and $\hat{n}$. We boost the momenta of the top-quarks into the zero-momentum frame (ZMF) of the di-top system and define the unit vector $\hat{k}$ in the direction of the top quark, which implies a direction $-\hat{k}$ for the anti-top quark. The four-momenta of the leptons are also boosted into the ZMF frame of the $t \bar{t}$ system and subsequently to the ZMF frame of the parent top-quark. We denote their directions of flight as $\hat{\ell}^+$ and $\hat{\ell^-}$. The scattering angle of the top quark with respect to the direction of flight of one of the protons $\hat{p}$ is then obtained by $\cos \theta_t = \hat{p} \cdot \hat{k}$, and an orthonormal unit vector can be defined via 
\begin{equation}
    \hat{n} = \frac{\textrm{sign}{(\cos\theta_t})}{\sin\theta_t} (\hat{p} \times \hat{k}) \;.
\end{equation}
The last required coordinate can be then defined as $\hat{r} = - \hat{n} \times \hat{k}$. 
We define the angles with the lepton's direction of flight  as $\cos \theta_{\hat{a}}^\pm = \pm \ell^\pm \cdot \hat{a}$ with $\hat{a} \in \{\hat{k}, \hat{r}, \hat{n}\}$. The $t\bar{t}$ cross section after integrating over azimuthal angles can be then parametrised as~\cite{Bernreuther:2004jv,Aguilar-Saavedra:2022uye,Rubenach:2023opp,CMS:2019pzc}
\begin{equation}
    \frac{1}{\sigma}\frac{d \sigma}{d \cos\theta^+_{\hat{a}} d \cos\theta^+_{\hat{b}}} = \frac{1}{4} ( 1 + B^+_{\hat{a}} \cos\theta^+_{\hat{a}} + B^-_{\hat{a}} \cos\theta^-_{\hat{a}} - C_{\hat{a}\hat{b}} \cos\theta^+_{\hat{a}} \cos\theta^-_{\hat{b}}) \;,
\end{equation}
where $\hat{a}, \hat{b} \in \{ \hat{k}, \hat{r}, \hat{n} \}$. The vector $\vec{B}^\pm$ contains details about the polarisations of the top quarks, while the matrix $C$ captures information regarding the spin correlation between the two quarks. This implies that the quantity $\cos \theta^{+}_{\hat{a}} \cos \theta^{-}_{\hat{b}}$ is sensitive to information regarding the spin correlations, and one can construct appropriate observables for discriminating between signal and background for the $t\bar{t}$ system. The observable 
\begin{equation}
    \label{eq:chel}
    \chel = - \cos\theta^+_{\hat{k}} \cos\theta^-_{\hat{k}} - \cos\theta^+_{\hat{r}} \cos\theta^-_{\hat{r}} - \cos\theta^+_{\hat{n}} \cos\theta^-_{\hat{n}} \;,
\end{equation}
is utilised by CMS and ATLAS in analyses of the $t \bar t$ final state.

In order to calculate $\chel$ distributions for different \cp-mixed scalars, we consider the invariant mass distribution of the di-top final state $m_{t \bar{t}}$ in different $\chel$ regions, in accordance with the bin definitions of Ref.~\cite{CMS:2019pzc}. We additionally take the expected SM background contributions as simulated by CMS in Ref.~\cite{CMS:2019pzc}, which already encode the higher-order corrections applied in that analysis, including electroweak effects and QCD higher-order corrections as used by CMS. The background shapes and rates in our analysis therefore implicitly include these corrections. For the signal contributions, we have simulated events with \madgraph.

We can write the full $g g \rightarrow t \bar{t}$ amplitude ${\cal{M}}$ as
\begin{align}
	\lvert{\cal{M}}\rvert^2   ={}& \lvert{\cal{M}}_\text{QCD}\rvert^2+
        \lvert{\cal{M}}_{h_1}\rvert^2 + \lvert{\cal{M}}_{h_2}\rvert^2  \nonumber\\
        &
		+ 2\, \mathrm{Re}\left({\cal{M}}_{h_1} {\cal{M}}_\text{QCD}^{\scaleobj{1.2}{*}}\right) 
		+ 2\, \mathrm{Re}\left({\cal{M}}_{h_2} {\cal{M}}_\text{QCD}^{\scaleobj{1.2}{*}}\right)
		+ 2\, \mathrm{Re}\left({\cal{M}}_{h_1} {\cal{M}}_{h_2}^{\scaleobj{1.2}{*}}\right) \;,
	\label{eq:amp_contrs}
\end{align}
separating the individual resonant, signal--background interference and signal--signal interference contributions. The SM background contributions from $g g \rightarrow t \bar{t}$ arising from QCD are labelled as ${\cal{M}}_\text{QCD}$\,, and ${\cal{M}}_{h_i}$ (for $i = 1,2$) denotes the amplitude arising from diagrams involving $h_i$. We produce events from each contribution separately by appropriately modifying the matrix element of \madgraph, which enables us to rescale each contribution with appropriate K~factors to take into account NLO effects. For the QCD contributions we use a K~factor of $K_\text{QCD} = 1.6$~\cite{Catani:2019,Kidonakis:2022}, while for the scalar contributions we obtain $K_{h_i}$ by calculating the $gg \rightarrow h_i$ prediction with \sushi~\cite{Harlander:2012pb,Harlander:2016hcx} through the \higgstools framework~\cite{Bahl:2022igd} and dividing by the LO prediction by \madgraph. Resonant contributions are rescaled by their respective K~factors, and interference contributions are rescaled by the square root of the product of the relevant K~factors, e.g.~the signal--signal interference is rescaled by $\sqrt{K_{h_1} K_{h_2}}$. 

We perform our analysis at parton-level, requiring that all leptons have a transverse momentum $p_T(\ell) > 20$~GeV and pseudorapidity $\lvert \eta (\ell) \rvert < 2.5$. At least one lepton must have a transverse momentum greater than $25$~GeV. Similarly, jets must satisfy $p_T(j) > 30$~GeV and $\lvert \eta (j) \rvert < 2.4$. We use the Monte-Carlo-truth information to identify the four-momenta of the tops and compute $m_{t \bar{t}}$ as well as $\chel$. We thus obtain histograms for $m_{t \bar{t}}$ in different $\chel$ bins for the signal contributions, using the same bins as CMS. In order to account for detector resolution effects that are absent at parton level, we smear the simulated $m_{t \bar{t}}$ distribution with a Gaussian of width $\SI{8}{\percent}$, where the value is chosen such that the expected significance for benchmark scenarios reproduces the CMS result.
\begin{figure}
    \centering
    \includegraphics[width=1.0\textwidth]{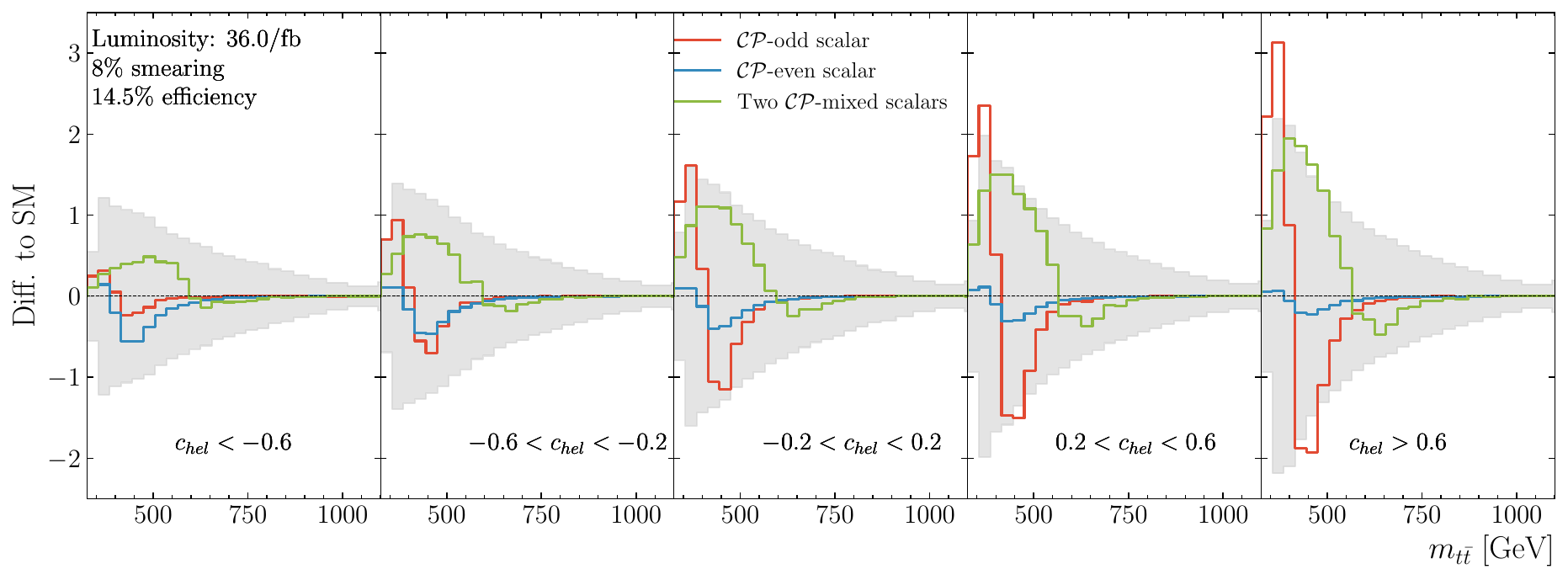}
    \caption{Difference between the simulated BSM contributions and the SM background prediction for the invariant mass distribution of the $t\bar{t}$ system, shown for different bins of the spin–correlation variable $\chel$. Results are displayed for a \cp-odd scalar (red), a \cp-even scalar (blue), and two \cp-mixed scalars (green). The gray bands indicate the systematic uncertainties reported by CMS~\cite{CMS:2019pzc}. An integrated luminosity of $36~\mathrm{fb}^{-1}$ at $\sqrt{s} = 13~\mathrm{TeV}$ and an 8\% smearing of $m_{t\bar{t}}$ are applied.}
    \label{fig:mtt_chel_plot}
\end{figure}
 
We show the BSM contributions to the $m_{t\bar{t}}$ distribution in different $\chel$ bins in~\cref{fig:mtt_chel_plot} for certain scenarios, as well as the systematic uncertainties reported by CMS~\cite{CMS:2019pzc} as gray bands. The displayed lines for the $\mathcal{CP}$-even and -odd scalars correspond to a single scalar with couplings $c_{t,1}=0.9$ and $\tilde{c}_{t,1} = 0.9$, respectively. The mass and width are $M_{h_1} = 400\,\text{GeV}$ and $\Gamma_{h_1} = 16\,\text{GeV}$ for both cases. For the scenario with two $\mathcal{CP}$-mixed scalars we use the parameters
\begin{align}
    \label{eq:benchmark_masscontributions}
    &c_{t,1}=0.7\;,\quad \tilde c_{t,1}=-1\;,\quad c_{t,2}=-0.8\;,\quad \tilde c_{t,2}=-1.1\;,\nonumber\\
    &M_{h_1} = \SI{572.81}{\giga\electronvolt}\;,\quad M_{h_2} = \SI{593.07}{\giga\electronvolt}\;, \nonumber\\
    &\Gamma_{h_1} = \SI{56.64}{\giga\electronvolt}\;,\quad\Gamma_{h_2} = \SI{20.1}{\giga\electronvolt}\;, \\
    &\bm{{Z}} = \begin{pmatrix}
        0.63-0.06 i & 0.76-0.01 i \\
        -0.75+0.01 i & 0.64-0.04 i
    \end{pmatrix} \;,\nonumber
\end{align}

which can be mapped to the c2HDM.\footnote{See~\ccite{Bahl:2025you} for the mapping between the simplified model used in this work and the c2HDM parameters.} We note that we have applied an efficiency factor of $14.5\%$ selected such that our expected significance for various single scalar hypotheses matches the corresponding numbers provided by CMS. The total number of events $n$ in each bin for a particular signal hypothesis can then be obtained by adding our histograms with the signal contributions $s$ to the histograms for the expected SM background $b$ as simulated by CMS in~Ref.~\cite{CMS:2019pzc}. We additionally take into account the systematic uncertainties $\sigma_b$ taken from CMS. The significance of the $m_{t\bar{t}}$ bin $i$ for each $\chel$ bin is then calculated as
\begin{align}
    \label{eq:significance}
    Z(s,b,\sigma_{b})_{i} =  \sqrt{2\left[ n \ln\!\left(\frac{n(b+\sigma_{b}^2)}{b^2+n\sigma_{b}^2}\right) - \frac{b^2}{\sigma_{b}^2}\ln\!\left(1+\frac{\sigma_{b}^2(n-b)}{b(b+\sigma_{b}^2)}\right) \right]} \;.
\end{align}
We subsequently combine the significances from different $m_{t \bar t}$ bins by summing in quadrature, 
\begin{align}
    Z = \sqrt{\sum_i \left(Z(s,b,\sigma_b)_i\right)^2} \;.
\end{align}
\begin{figure}[!ht]
    \centering
    \includegraphics[width=0.6\textwidth]{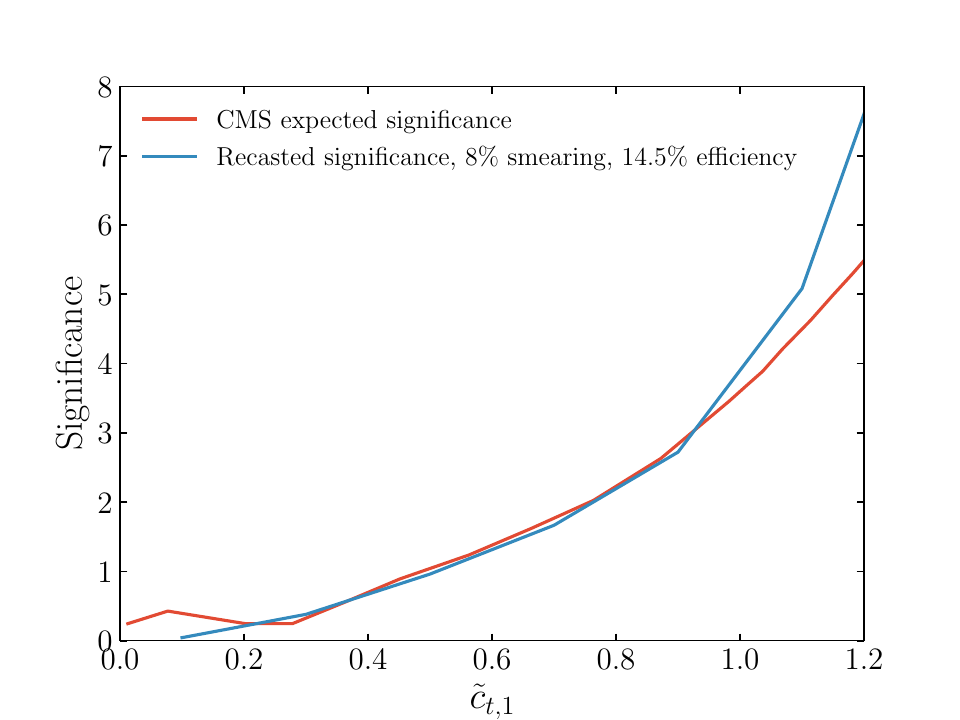}
    \caption{The calculated recast significance is shown in blue, with $8\%$ smearing and the efficiency set to $14.5\%$, in comparison with the CMS expected significance which is shown in red.}
    \label{fig:cms_comparison}
\end{figure}

In~\cref{fig:cms_comparison} we show a comparison of our calculated significance for the case of a single \cp-odd scalar with mass of $400\,\text{GeV}$ and the expected significance of CMS in their analysis for a \cp-odd scalar. For significance values $Z \lesssim 5$ our analysis is able to reproduce the CMS significance rather well. We therefore expect that our obtained significances for more complicated scenarios with more than one scalar will be indicative of the expected significances in actual LHC experiments. 

\subsection{Searches for new scalar particles in four-top final states}
\label{sec:tttt}

The production process of four top quarks has been observed by both the ATLAS~\cite{ATLAS:2023ajo} and CMS~\cite{CMS:2023ftu} collaborations only recently. Processes with three or four top quarks in the final state typically have lower production cross-sections than those with two top quarks and are experimentally demanding due to the large object multiplicity and the challenge of reconstructing the top quarks. In searches for BSM resonances produced in association with top quarks an additional challenge is to correctly reconstruct and identify the top quarks from the resonance decay. Although the rarity of the signal process results in inferior sensitivity of the four-top channel compared to the di-top channel, it can provide complementary information. Searches for top-philic resonances are performed by the ATLAS~\cite{ATLAS:2022rws,ATLAS:2024jja,ATLAS:2024bsm} and CMS collaborations~\cite{CMS-TOP-18-003}.

Similarly as in the previous section, we proceed by estimating the sensitivity of LHC searches in the four-top-quark final state to the interference effects by reinterpreting a search for four-top-quark production performed by the CMS collaboration using \SI{137}{\per\femto\barn} of $13\tev$ \(pp\) collision data~\cite{CMS-TOP-18-003}. We focus on the four-top-quark process \(p p \to h_{1,2} t \bar{t} \to t \bar{t} t \bar{t}\) and the three-top-quark processes \(p p \to h_{1,2} t W^{-} \to t \bar{t} t W^{-}\) and \(p p \to h_{1,2} t \bar{q} \to t \bar{t} t \bar{q}\) (and similarly for an initially produced anti-top quark). The analysis targets final states with two leptons of the same electric charge, as most large SM background processes like \(t\bar{t}\) or \(Z\) + jets produce opposite-sign leptons. Additionally, the large jet multiplicity and the presence of large hadronic activity \(H_{\text{T}}\), defined as the scalar sum of the transverse momenta of all jets, allows a discrimination of the signal process from its backgrounds. Events of the signal process are generated following a similar approach using MC event generators as for the di-top analysis. In contrast to the di-top analysis we exploit the fact that a \madanalysis~\cite{Conte:2018vmg} implementation of the analysis of Ref.~\cite{Darme:2020ma5} has been released. Therefore, we simulate events with \madgraph and \pythia, using \delphes~\cite{Delphes2014} for the parametrisation of the CMS detector response. The resonant signal contributions are rescaled by the product of the relevant K~factors,\footnote{The K-factors are obtained by comparing for each mass the obtained LO cross-sections with the top-associated cross-section values recommended by the LHC Higgs working group~\cite{LHCHiggsCrossSectionWorkingGroup:2016ypw}.} for instance for a \SI{500}{\giga\electronvolt} scalar produced in association with a top-quark pair, the K~factor is \num{1.25}, while for the production in association with a single top quark it is \num{1.32}.

The selected events of the analysis are required to contain two same-sign charged leptons (electrons or muons) or at least three leptons and jets. The main observables to construct signal regions enriched in respective four-top or three-top signal processes are the lepton multiplicity \(N_{l}\), the multiplicity of jets \(N_{j}\), and the multiplicity of \(b\)-jets (jets originating from a \(b\)-quark) \(N_{b}\). Leptons which satisfy requirements on their transverse momentum \(p_{\text{T}} > \SI{20}{\giga\electronvolt}\) and pseudo-rapidity \(\abs{\eta} < 2.4\) (for muons: \(\abs{\eta} < 2.5\)) are considered when determining \(N_l\). Jets have to satisfy \(p_{\text{T}} > \SI{40}{\giga\electronvolt}\) and \(\abs{\eta} < 2.4\) to be considered for \(N_{j}\), while \(b\)-jets have to satisfy \(p_{\text{T}} > \SI{25}{\giga\electronvolt}\), \(\abs{\eta} < 2.5\) and a requirement on a multivariate discriminant when determining \(N_{b}\). All selected events have to pass the requirements of a baseline selection which requires \(H_{\text{T}} > \SI{300}{\giga\electronvolt}\), missing transverse momentum \(p_{\text{T}}^{\text{miss}} > \SI{50}{\giga\electronvolt}\), more than two jets and more than two \(b\)-jets. Further requirements include a leading lepton with \(p_{\text{T}} > \SI{25}{\giga\electronvolt}\) and a same-charge trailing lepton with \(p_{\text{T}} > \SI{20}{\giga\electronvolt}\). Same-sign electron pairs with an invariant mass below \SI{12}{\giga\electronvolt} are rejected to suppress low-mass resonance backgrounds. Additionally, events with a third lepton (\(p_{\text{T}} > \SI{7}{\giga\electronvolt}\) for electrons, \(p_{\text{T}} > \SI{5}{\giga\electronvolt}\) for muons) forming an opposite-sign same-flavour pair with invariant mass below \SI{12}{\giga\electronvolt} or between \SIrange{76}{106}{\giga\electronvolt} are excluded. Events that pass the baseline selection are classified in one of the 14 signal regions based on \(N_l\), \(N_b\) and \(N_j\).

The expected significance of the signals \(Z(s,b,\sigma)\) is estimated for each defined signal region \(i\) of the analysis with \(s\) signal events  and \(b\) background events with an uncertainty \(\sigma_{b}\) in that region with the same approach as for the di-top analysis, using~\cref{eq:significance}. In the following, the largest significance  value \(\max_{r \in \text{regions}} Z(s,b,\sigma_{b})_{r}\) among all signal regions is reported. We evaluate the significance and selection efficiency for a single scalar with mass \(M_{h_1} = \SI{500}{\giga\electronvolt}\) and width of \(\Gamma_{h_1} = \SI{10}{\giga\electronvolt}\). The total signal efficiency is \SI{0.87}{\percent}.  The corresponding largest signal efficiency in a single signal region (SR) is \SI{0.19}{\percent} in SR8 (\(N_l = 2\), \(N_b \geq 4\), \(N_j \geq 5\)), followed by \SI{0.13}{\percent} in SR4 (\(N_l = 2\), \(N_b = 3\), \(N_j = 5\)). 
\begin{figure}
    \centering
    \includegraphics[width=0.4\textwidth]{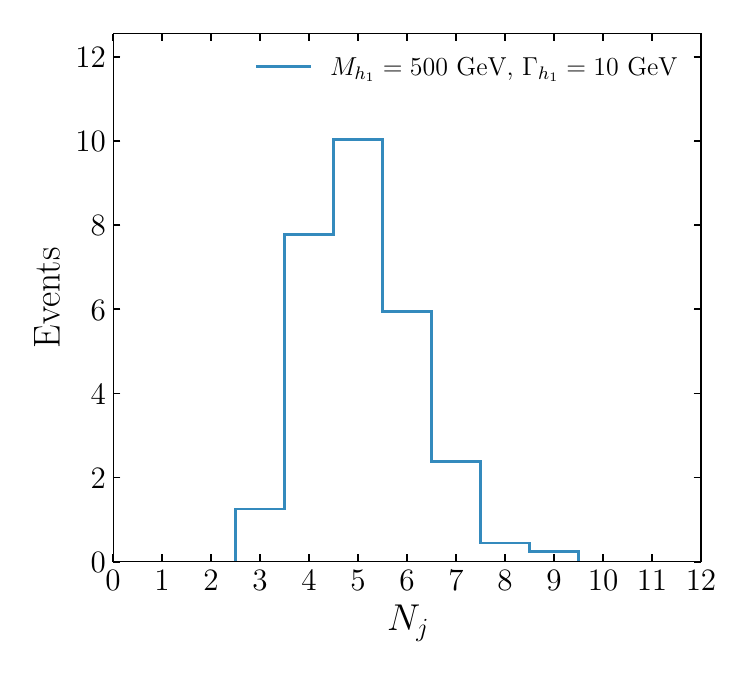}
    \quad
    \includegraphics[width=0.4\textwidth]{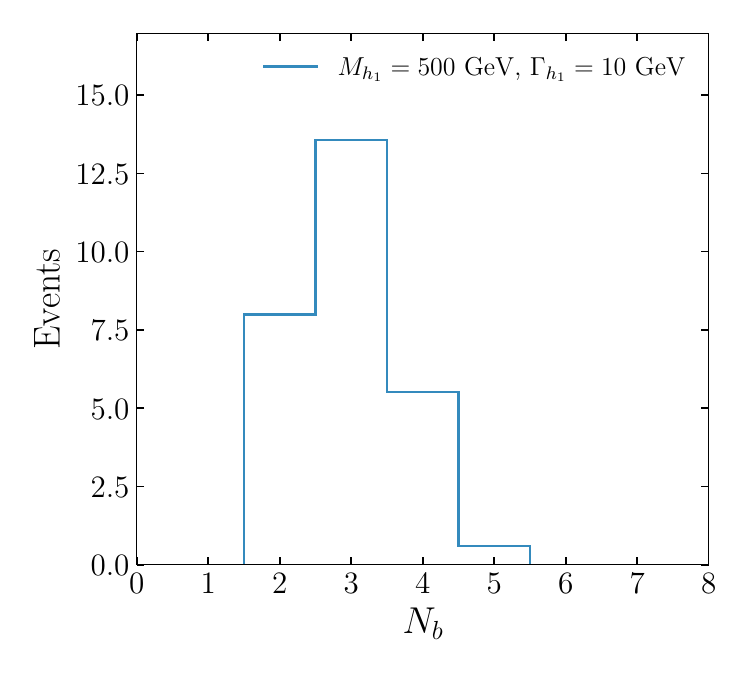}
    \caption{Distributions of jet multiplicity \(N_j\) (left) and \(b\)-jet multiplicity \(N_b\) (right) in the analysis region of the CMS four-top analysis implemented in \madanalysis for a signal with a single scalar which has a mass of \(M_{h_1} = \SI{500}{\giga\electronvolt}\) and a width of \(\Gamma_{h_1} = \SI{10}{\giga\electronvolt}\). The distributions are scaled to the theory cross-section and a luminosity of $137/\text{fb}$.}
    \label{fig:4top-plots-scalar500}
\end{figure}

\Cref{fig:4top-plots-scalar500} shows the inclusive jet and \(b\)-jet multiplicity distributions, illustrating that signal events are characterised by a large jet multiplicity and 2--4 \(b\)-jets, which is why SR8 and SR4 are most sensitive to four-top signals. The amount of initial generated events and the amount of events passing the event selection after each respective requirement is shown for SR8 in~\cref{tab:cutflow_500GeV}. 

\begin{table}[!ht]
  \centering
  \caption{Cut-flow of selected events in the analysis region of the CMS four-top analysis implemented in \madanalysis in the signal region with the largest signal efficiency, SR8, for a signal with a single scalar which has a mass of \(M_{h_1} = \SI{500}{\giga\electronvolt}\) and a width of \(\Gamma_{h_1} = \SI{10}{\giga\electronvolt}\).}
  \label{tab:cutflow_500GeV}
  \begin{tabular}{lcccccccc}
    \toprule
    Cut & Initial & Baseline & $m_{ll} >$  12 GeV & Veto 3l & SR & 2 leptons & $\geq$4 bjets & $\geq$5 jets \\
    \midrule
    Events & 99938 & 1579 & 1576 & 1514 & 868 & 716 & 165 & 165 \\
    \bottomrule
  \end{tabular}
\end{table}

\section{Interference patterns in di-top and four-top production}
\label{sec:top_interference}

With the analysis setups for both di-top and four-top production in place, we now turn to the study of interference patterns in these processes and assess their impact on the discovery sensitivity for BSM scalar resonances.

\begin{figure}
    \centering
    \includegraphics[height=0.38\linewidth]{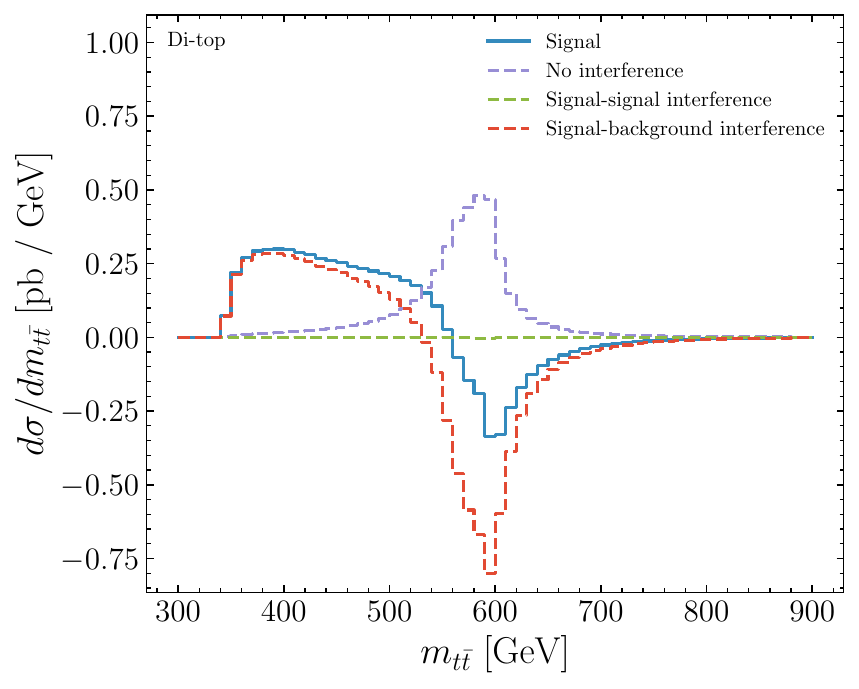}
    \includegraphics[height=0.38\linewidth]{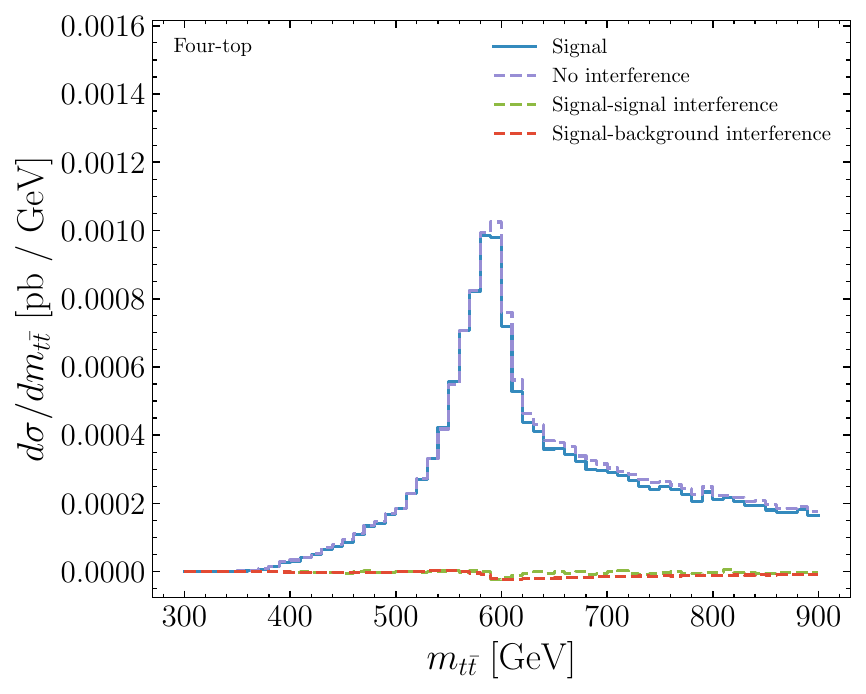}
    \caption{Parton-level invariant mass spectra (including the folding with the gluon parton distribution functions) for di-top production (left) and four-top production (right) for the benchmark point specified in~\cref{eq:benchmark_masscontributions}. The purple line labelled ``No interference'' shows the incoherent sum of the two resonant signal contributions. The green line shows the contribution from signal–signal interference, the red line shows the contribution from signal–background interference, and the blue dashed line labelled ``Signal'' shows the full signal prediction including both interference terms.}
    \label{fig:mass-spectra}
\end{figure}

We examine the parton-level mass spectra for both di-top and four-top production. \Cref{fig:mass-spectra} displays the parton-level invariant mass spectra without accounting for detector smearing for the different signal contributions: the left panel corresponds to the di-top process, while the right panel shows the four-top process. For the latter, the two top quarks leading in transverse momentum at parton level are considered. As a benchmark point, we use here the parameter point of~\cref{eq:benchmark_masscontributions} which can be realised in the c2HDM and illustrates the difference in interference patterns between the two channels.

In the di-top panel, the purple curve labelled ``No interference'' shows the incoherent sum of the two resonant signal contributions, that is, the result obtained if interference among the two scalars and with the SM background is ignored. The green curve shows the contribution from signal--signal interference. For this example scenario, it is a negligible contribution, which is negative in the region where the two resonances overlap and therefore slightly reduces the rate on the high-mass shoulder. The red curve shows the signal--background interference contribution, which generates the characteristic peak--dip structure around $m_{t\bar t} \simeq 400$–\SI{420}{\giga\electronvolt}. The blue dashed curve labelled ``Signal'' denotes the full signal prediction, obtained by combining the incoherent resonant contribution with both types of interference terms. Above the resonance region this spectrum approaches the incoherent result, while above the $t \bar t$ threshold it develops a pronounced peak followed by a dip in the vicinity of the poles.

In the four-top process, shown in the right panel, the SM background is much smaller than in the di-top case. This is reflected in the red curve, which shows that the signal–background interference term remains very small over the whole mass range. For this example scenario, the signal--signal interference contribution (green) is also subdominant compared to the resonant contribution. Consequently, the dominant new-physics contribution arises from resonant production followed by on-shell decays. As a result, the full signal prediction (blue) is almost identical to the incoherent sum of the two resonances (purple dashed) and has a lineshape that closely resembles a Breit–Wigner peak close to the physical pole masses. The peak is slightly broader than the no-interference contribution in the di-top case because $m_{t\bar t}$ is computed from the two leading top quarks in transverse momentum rather than from a single well-defined top--anti-top pair.

This comparison illustrates the complementarity of the two channels. Di-top production yields a larger signal rate, but its spectrum is very sensitive to interference effects and can exhibit pronounced peak–dip structures. In contrast, the four-top final state provides a much cleaner resonance signal that can directly be associated with the underlying poles of the resonantly produced particles.

\begin{figure}
    \centering
    \includegraphics[height=0.38\linewidth]{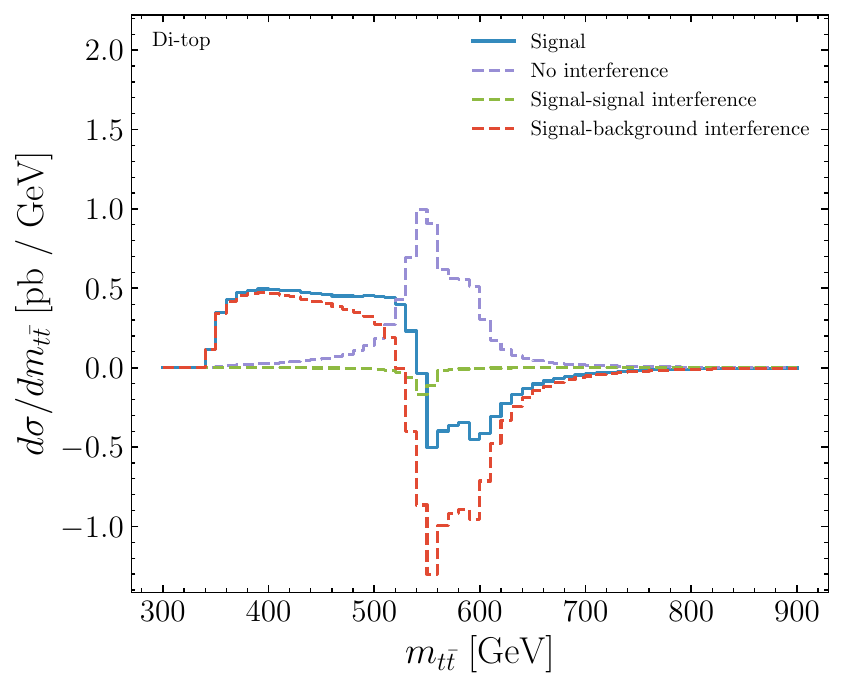}
    \includegraphics[height=0.38\linewidth]{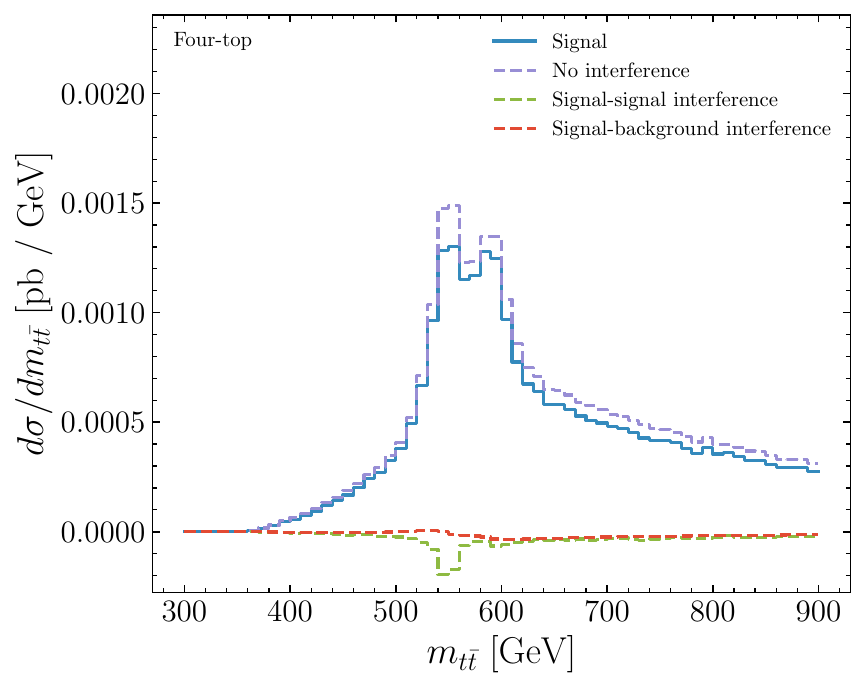}
    \caption{Parton-level invariant mass spectra (including the folding with the gluon parton distribution functions) for di-top production (left) and four-top production (right) for the \cp-mixed scenario~1 with the tree-level masses fixed to  $m_{h_{1,\mathrm{tree}}} = m_{h_{2,\mathrm{tree}}} = 550\ \mathrm{GeV}$. The couplings are fixed to $c_{t,1} = 0.8$, $\tilde{c}_{t,1} = 0.2$, $c_{t,2} = 0.6$, and $\tilde{c}_{t,2} = 1$. The purple line labelled ``No interference'' shows the incoherent sum of the two resonant signal contributions. The green line shows the contribution from signal--signal interference, the red line shows the contribution from signal--background interference, and the blue dashed line labelled ``Signal'' shows the full signal prediction including both interference terms.}
    \label{fig:mass-spectra-2}
\end{figure}

We now evaluate the effects of these interference patterns on a physics analysis. To this end, we define two benchmark scenarios, each corresponding to states with mixed \cp properties but probing different mass regimes:

\begin{itemize}
    \item \textbf{Case~1.} We fix the couplings to $c_{t,1} = 0.8$, $\tilde{c}_{t,1} = 0.2$, $c_{t,2} = 0.6$, and $\tilde{c}_{t,2} = 1$. The tree-level mass of the first state is set to $m_{h_{1,\mathrm{tree}}} = 550\ \mathrm{GeV}$, while the second tree-level mass parameter is varied in the range $m_{h_{2,\mathrm{tree}}} \in [350, 1000]\ \mathrm{GeV}$. The decay widths are calculated to be $\Gamma_{h_1} \simeq 10\ \mathrm{GeV}$ and $\Gamma_{h_2} \in [4, 70]\ \mathrm{GeV}$. In addition to the mass scan, we perform a dedicated scan in $\tilde{c}_{t,2}$, keeping both tree-level masses fixed to $m_{h_{1,\mathrm{tree}}} = m_{h_{2,\mathrm{tree}}} = 550\ \mathrm{GeV}$.
    
    \item \textbf{Case~2.} We fix the couplings to $c_{t,1} = -0.4$, $\tilde{c}_{t,1} = 0.5$, $c_{t,2} = 0.4$, and $\tilde{c}_{t,2} = 0.8$. Here, the first tree-level mass is set to $m_{h_{1,\mathrm{tree}}} = 750\ \mathrm{GeV}$, while the second tree-level mass is varied in the range $m_{h_{2,\mathrm{tree}}} \in [350, 1200]\ \mathrm{GeV}$. The decay widths are calculated to be $\Gamma_{h_1} \simeq 15\ \mathrm{GeV}$ and $\Gamma_{h_2} \in [3, 50]\ \mathrm{GeV}$. As in case~1, we also perform a scan in $\tilde{c}_{t,2}$ with both masses fixed to $m_{h_{1,\mathrm{tree}}} = m_{h_{2,\mathrm{tree}}} = 550\ \mathrm{GeV}$.    
\end{itemize}
For both scenarios, the complex $\bm{Z}$-matrix elements are calculated point by point as discussed above. The two scenarios are chosen to allow for large interference contributions (see~\ccite{Bahl:2025you} for analytic expressions for the di-top case). \Cref{fig:mass-spectra-2} shows the di- and four-top interference pattern in case~1 for the exemplary point of $m_{h_{1,\mathrm{tree}}} = m_{h_{2,\mathrm{tree}}} = 550\ \mathrm{GeV}$. For the di-top final-state, we observe a large signal--background interference. For both the di- and four-top final states, the signal--signal interference is relatively small. Larger effects occur if the loop-level masses are closer to each other.

\begin{figure}
    \centering
    \includegraphics[width=0.6\textwidth]{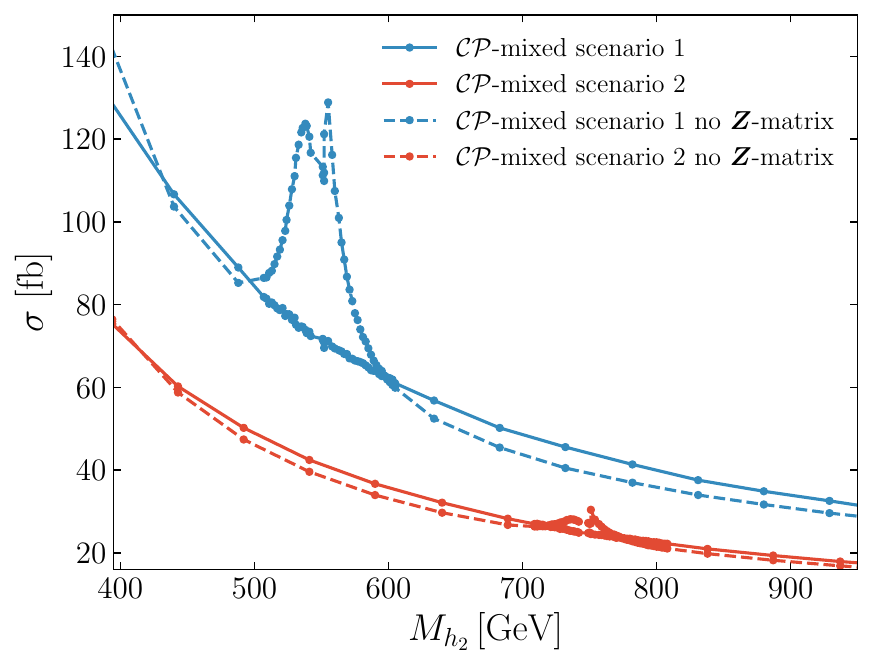}
    \caption{The di-top cross section as a function of the mass of the second scalar $M_{h_2}$ is shown with and without taking into account the contribution of the $\bm{Z}$-matrix. The sampling density is increased in the region of $M_{h_1}\simeq M_{h_2}$.}
    \label{fig:scenarios_mass_xsec}
\end{figure}

\begin{figure}[ht!]
    \centering    
    \includegraphics[width=0.47\textwidth]{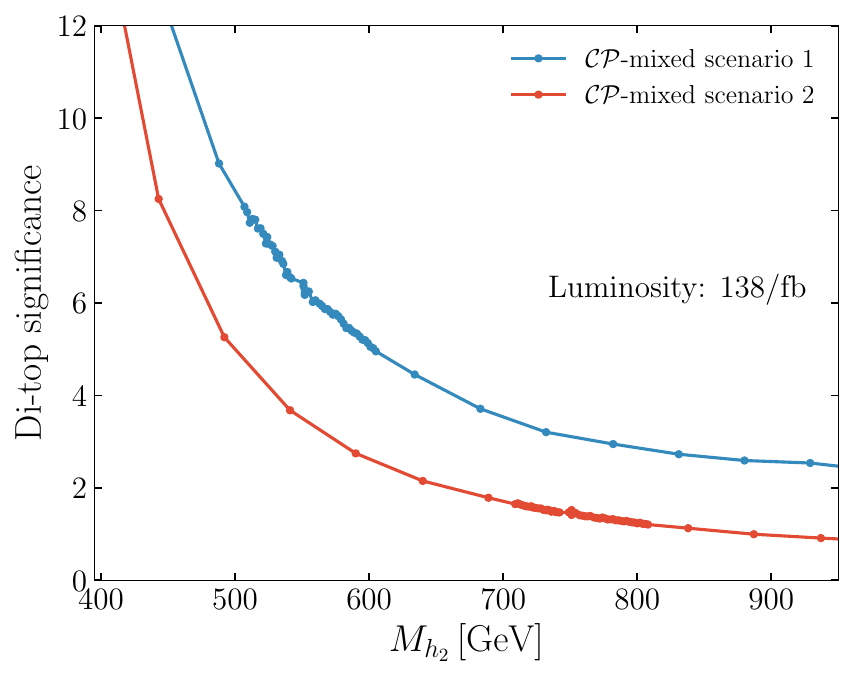}
    \includegraphics[width=0.46\textwidth]{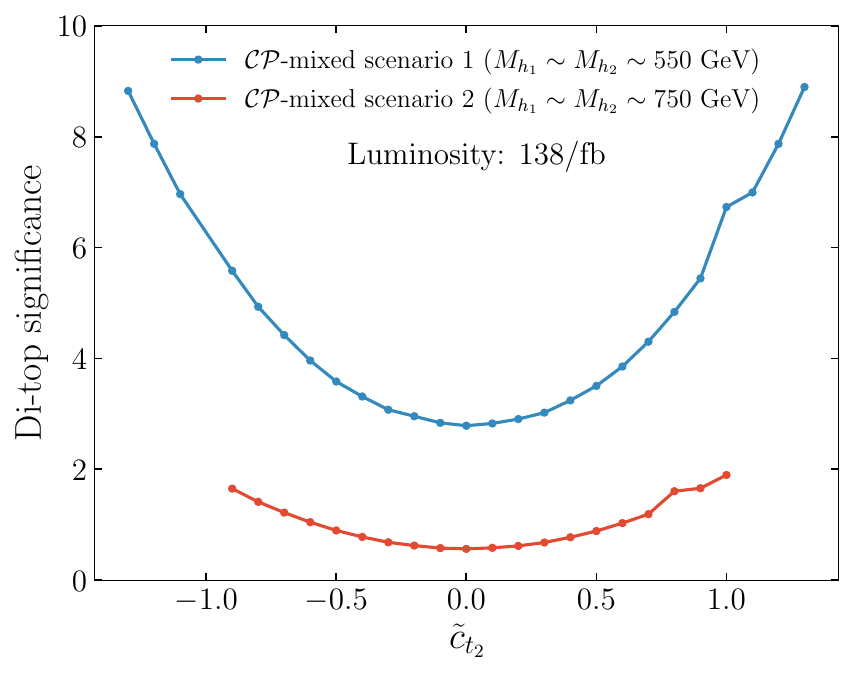}
    \caption{On the left, we show the di-top discovery significance as a function of the mass of the second scalar, $M_{h_2}$, while the discovery significance of the di-top final state as a function of the \cp-odd coupling of the second scalar to the top-quark, $\tilde{c}_{t_2}$, is shown on the right. The $\bm{Z}$-matrix contributions are included for all results.}
    \label{fig:scenarios_mass_coupling_varied}
\end{figure}

For the evaluation of the discovery significance, we first focus on the di-top final state. To assess the impact of the intricate interference patterns on the experimental sensitivity, we show in~\cref{fig:scenarios_mass_xsec} the di-top cross section with and without including the $\bm{Z}$-matrix. For the case where the $\bm{Z}$-matrix is not taken into account, a resonant structure appears where the masses are degenerate. In this region where strong signal--signal interference occurs, a pronounced peak appears in the significance curve if loop-level mixing effects are ignored. This behaviour closely follows the observations made in the toy-model study discussed in~\cref{sec:toy_model}. It is therefore essential to account for loop-level mixing: ignoring it leads to a severe overestimation of the discovery significance or, in the absence of a signal, of the corresponding exclusion limits. If loop-level mixing is properly included, the significance curve becomes smooth, showing no visible imprint of signal--signal interference. This behaviour is again consistent with the toy-model results presented in~\cref{sec:toy_model}, once parton-shower and detector effects are taken into account, which effectively smear the parton-level invariant-mass distribution.

We additionally show the discovery significance as a function of $M_{h_2}$ for the two \cp-mixed benchmark scenarios under consideration in the left panel of~\cref{fig:scenarios_mass_coupling_varied} for an integrated luminosity of 138/fb. Following the behaviour of the total cross section, the significance decreases with increasing $M_{h_2}$ showing no sign of interference. The significance curves are not only smooth when varying the masses, but also when varying $\tilde c_{t_2}$, as shown in the right panel of~\cref{fig:scenarios_mass_coupling_varied}. Both benchmark scenarios exhibit a characteristic local minimum in the discovery significance for $\tilde{c}_{t_2} = 0$. Apart from this feature, the dependence of the significance is symmetric with respect to the sign of the coupling.
\begin{figure}
    \centering
    \includegraphics[width=0.47\textwidth]{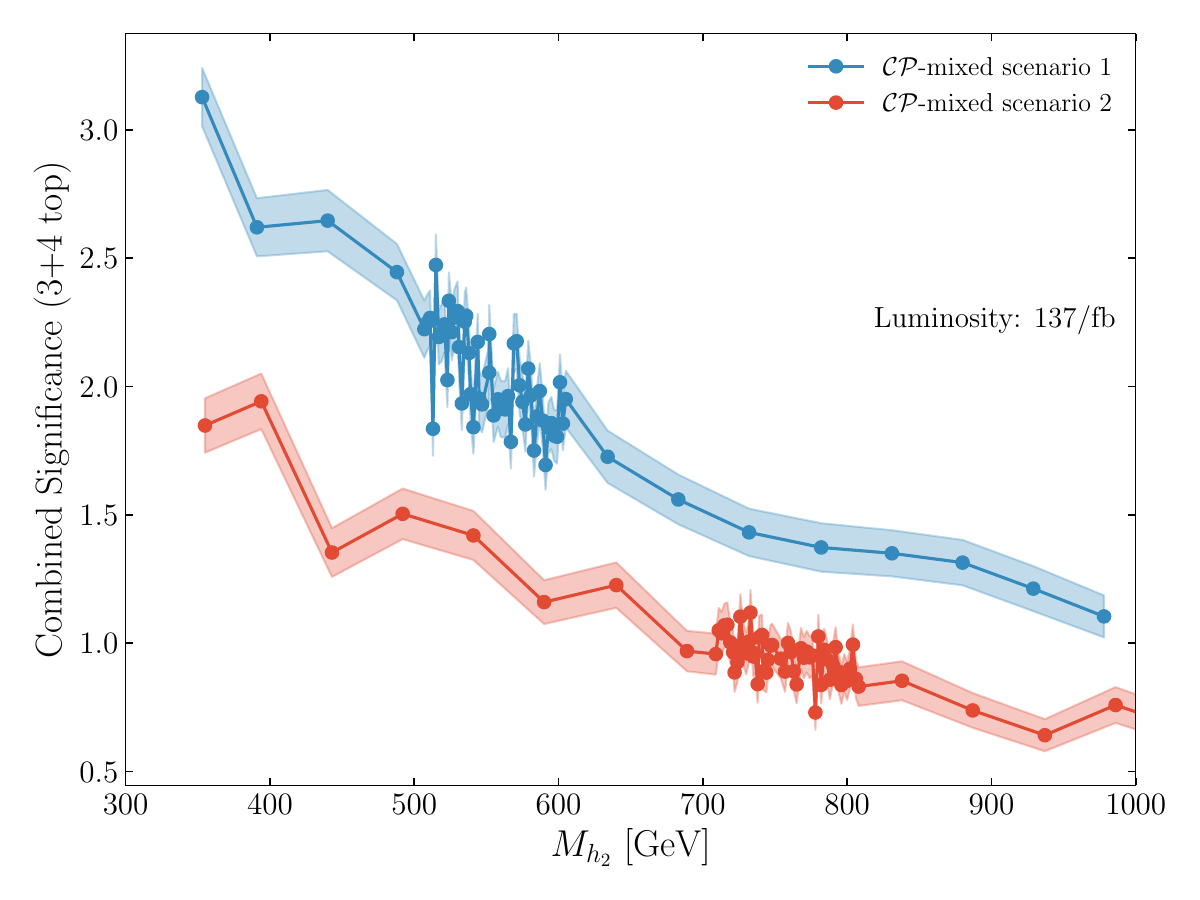}
    \includegraphics[width=0.47\textwidth]{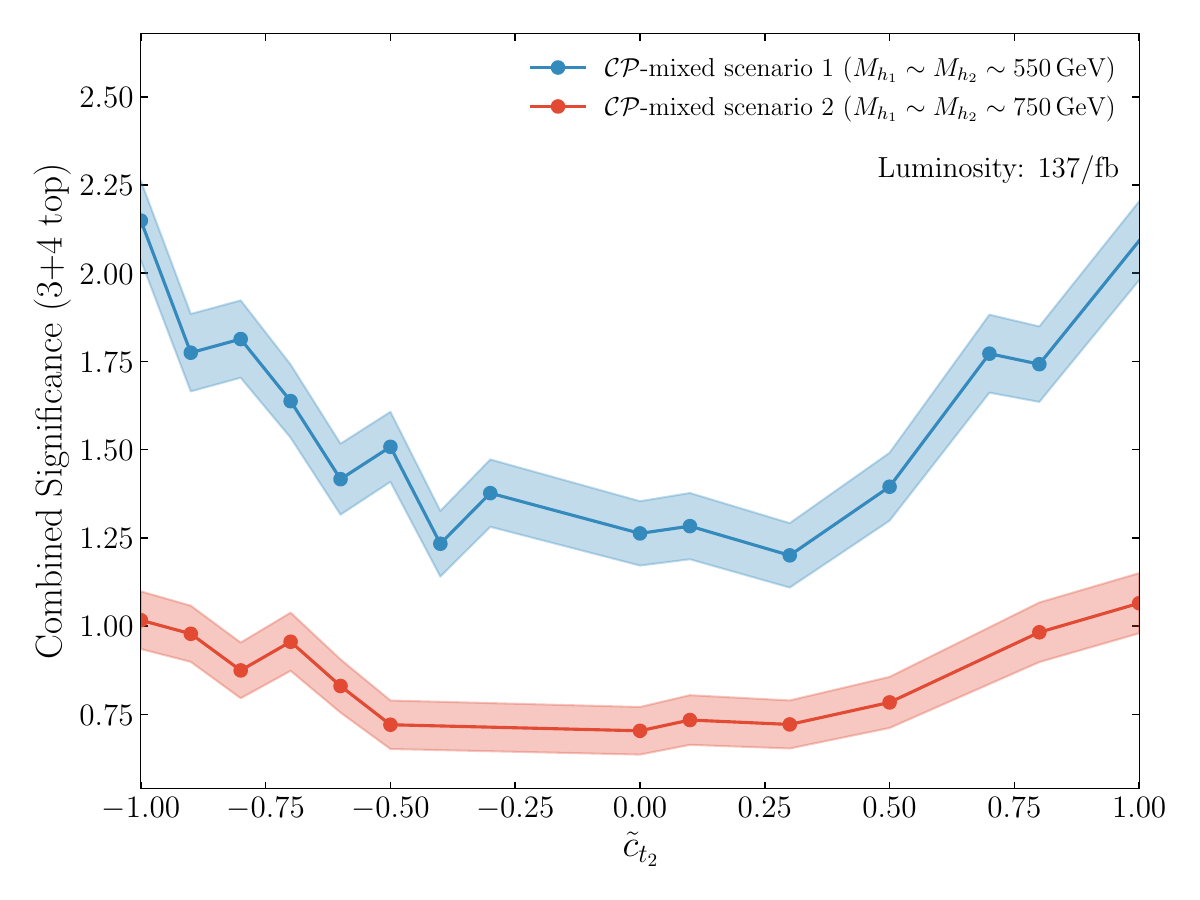}
    \caption{Four-top significance as a function of the mass of the second scalar $M_{h_2}$ (left) and its \cp-odd coupling to the top-quark $\tilde{c}_{t_2}$ (right). The shaded bands correspond to the estimate of the uncertainty on the significance using 1000 pseudo-experiments in which signal and background have been sampled from normal distributions. In the left panel, the sampling density is increased in the region of $M_{h_1}\simeq M_{h_2}$.}
    \label{fig:c2hdm-bp3-multitop-1}
\end{figure}

Next, we turn to the four-top final state, studying the same two benchmark scenarios. The left panel of~\cref{fig:c2hdm-bp3-multitop-1} shows the four-top discovery significance as a function of the second BSM mass for an integrated luminosity of 137/fb, in analogy to the left panel of~\cref{fig:scenarios_mass_coupling_varied}. In addition to the four-top processes, three-top processes have also been simulated and included, as the CMS analysis accounts for these as well. However, due to the significantly lower selection efficiency, their contribution to the significance is roughly an order of magnitude lower than that of the four-top process. Given the smaller overall rates compared to di-top production, we also show uncertainty bands. These are evaluated using a bootstrapping approach, in which 1000 toy experiments are performed, sampling signal and background yields from normal distributions with mean and standard deviation set to the expected yields and uncertainties, respectively.

As in the di-top case, the significance decreases with increasing $M_{h_2}$. The same qualitative features observed for di-top production are also present here: if loop-level mixing effects are ignored, a pronounced peak appears in the region of strong signal--signal interference as shown in~\cref{app:sig_scans}. For the case where loop-level mixing is correctly included, the significance becomes an almost smooth function of $M_{h_2}$ as shown in the left panel of~\cref{fig:c2hdm-bp3-multitop-1}.

The right panel of~\cref{fig:c2hdm-bp3-multitop-1} shows the significance as function of $\tilde{c}_{t_2}$. The relevant couplings are varied. The features closely resemble those observed in~\cref{fig:scenarios_mass_coupling_varied}, including the characteristic local minimum in the discovery significance for $\tilde{c}_{t_2} \approx 0$.

The fluctuations visible in both panels of~\cref{fig:c2hdm-bp3-multitop-1} are a consequence of only a small number of Monte-Carlo events ending up in SR8.

\section{Di-top and four-top interplay}
\label{sec:tt_tttt_interplay}

In the following, we study how information from the di-top and four-top channels can complement each other in scenarios with top-philic scalars, taking loop-level mixing effects into account. Summarising the discussion from the previous section, two robust features emerge. First, once loop-level mixing is included via the $\bm{Z}$-matrix, both channels exhibit smooth significance curves as masses or couplings are varied. The sharp artificial peaks that appear if loop-level mixing is ignored are absent, consistent with the behaviour observed in the toy-model study presented in~\cref{sec:toy_model}. Second, across all scanned points, we did not find regions where the di-top significance is small while the three- and four-top significances are large. The hierarchy is rather stable: the di-top selection is typically more sensitive for the same luminosity, while the channel with three and four tops in the final state provides complementary information, as in this case the $m_{t \bar t}$ distribution is less distorted by signal--background interference. The different interference patterns can be crucial for determining the underlying physics scenario, while the results obtained from the di-top channel alone may be difficult to interpret. Moreover, both channels can be combined in a simple and robust manner to enhance the overall sensitivity.

\subsection{Phenomenological examples in the c2HDM framework}

We further investigate the complementarity between the di-top and four-top channels within the framework of the softly broken \(\mathbb{Z}_2\)-symmetric complex two-Higgs-doublet model (c2HDM) \cite{Branco:2011iw, Gunion:2002zf, Barger:2009me}. Selected benchmark points in the model parameter space are used to illustrate specific features that may arise when interpreting searches in final states containing top quarks.

\begin{figure}
    \centering
    \includegraphics[height=0.8\linewidth,width=0.8\linewidth,keepaspectratio]{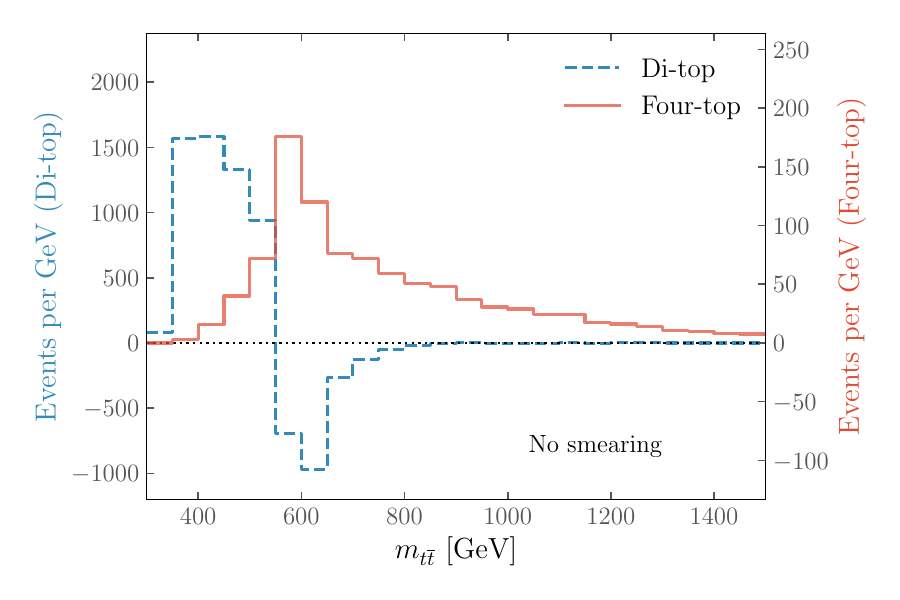}
    \caption{Comparison of the $m_{\ttb}$ distributions for a signal in the di-top and four-top final states with $M_{h_1} = \SI{572.81}{\giga\electronvolt}$ and $M_{h_2} = \SI{593.07}{\giga\electronvolt}$ for the benchmark point of~\cref{eq:benchmark_masscontributions}. The distributions are scaled to the respective cross-sections and a luminosity of \SI{137}{\per\femto\barn}.}
    \label{fig:c2hdm-bp3-4top-4}
\end{figure}

In order to illustrate that the channel comprising the three-top and four-top final states provides complementary sensitivity to the di-top channel,~\cref{fig:c2hdm-bp3-4top-4} shows the inclusive signal distributions for the di-top and four-top processes. In the displayed $m_{t \bar t}$ distributions for \(\ttb\) production in the c2HDM, using the parameter point of~\cref{eq:benchmark_masscontributions}, all interference effects are included, while the respective SM QCD backgrounds are not included. The blue dashed curve denotes the di-top signal (left-hand $y$ axis), while the red solid curve corresponds to the four-top signal (right-hand $y$ axis). Although it can be seen that the di-top signal yields a much larger event rate, it should be noted that this channel also suffers from substantially higher backgrounds. Moreover, the di-top distribution exhibits the characteristic peak–dip structure induced by interference with the background. The peak is not located at the physical mass values but --- as a consequence of the gluon parton-distribution function of the colliding protons --- starts at the $t \bar t$ threshold. While the dip of the distribution is correlated with the physical mass values, this information may be difficult to extract from the experimental result (see the discussion below). In contrast, the four-top signal displays a distinct, well-defined resonance peak at the location of the physical masses.

\begin{figure}
    \centering
    \includegraphics[height=0.8\linewidth,width=0.8\linewidth,keepaspectratio]{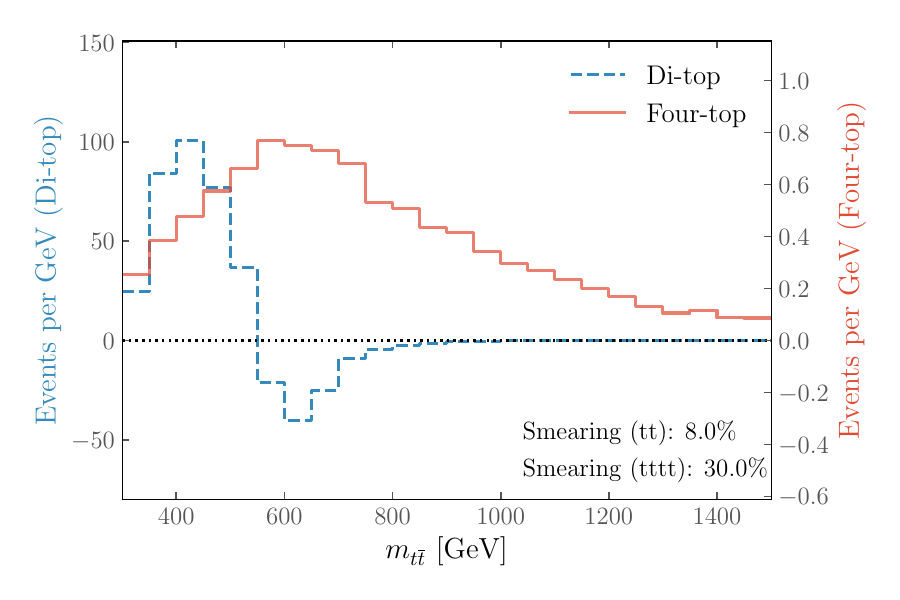}
    \caption{Comparison of the $m_{\ttb}$ distributions for a signal in the di-top and four-top final states with $M_{h_1} = \SI{572.81}{\giga\electronvolt}$ and $M_{h_2} = \SI{593.07}{\giga\electronvolt}$ for the benchmark point of~\cref{eq:benchmark_masscontributions}. The distributions are scaled to the respective cross-sections and a luminosity of \SI{137}{\per\femto\barn}. The finite mass resolution is considered by sampling the mass distribution with a Gaussian distribution with standard deviation of $8\%$ (di-top) and $30\%$ (four-top). In addition, the acceptance and efficiency of the two reference analyses are approximated by only considering a randomly selected subset of events corresponding to the product of acceptance and efficiency for the two searches, which are \SI{6.5}{\percent} for di-top and \SI{1}{\percent} for four-top.}
    \label{fig:c2hdm-bp3-4top-4_smear_acceptance}
\end{figure}

The distributions shown in~\cref{fig:c2hdm-bp3-4top-4} correspond to an idealised situation where no experimental smearing has been incorporated. If detector resolution and object-reconstruction effects are taken into account, the $m_{\ttb}$ distributions becomes broader. \Cref{fig:c2hdm-bp3-4top-4_smear_acceptance} shows the signals including both finite-resolution and acceptance/efficiency effects. The latter have been accounted for by applying a constant weight to each event, which only affects the total event yield. The finite resolution has been accounted for by Gaussian smearing of the mass distribution with the respective resolution taken from the experimental analyses. This has the consequence that for the di-top channel the dip is significantly less pronounced than for the case without experimental smearing. The remaining dominant feature is a distinct peak starting just above the $t \bar t$ threshold which is difficult to associate with the production of a particular BSM signal. In contrast, the four-top peak position remains stable and directly correlated to the physical masses of the produced particles, despite the much poorer mass resolution of \SI{30}{\percent} that is used here compared to the \SI{8}{\percent} resolution in the di-top channel. Note that here for simplicity we apply a flat efficiency factor, while in a realistic analysis the efficiency would vary across phase space. Even though the event numbers in the four-top case are small, experimental analysis can still be sensitive due to the very small SM background in specific signal regions (see e.g.~\cref{fig:c2hdm-bp3-multitop-1}).

\begin{figure}
    \centering
    \includegraphics[height=0.36\textwidth]{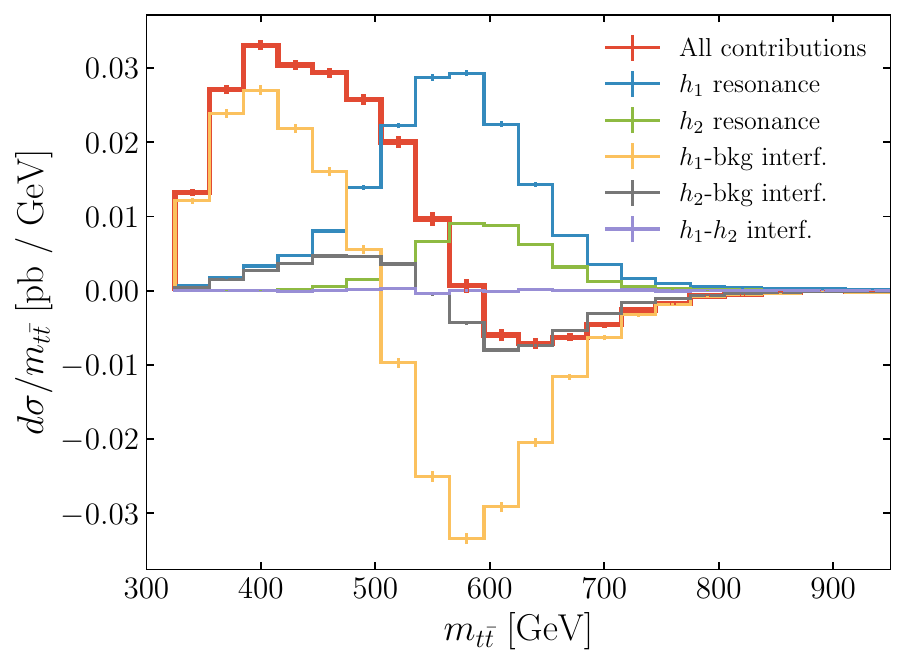}
    \includegraphics[height=0.36\textwidth]{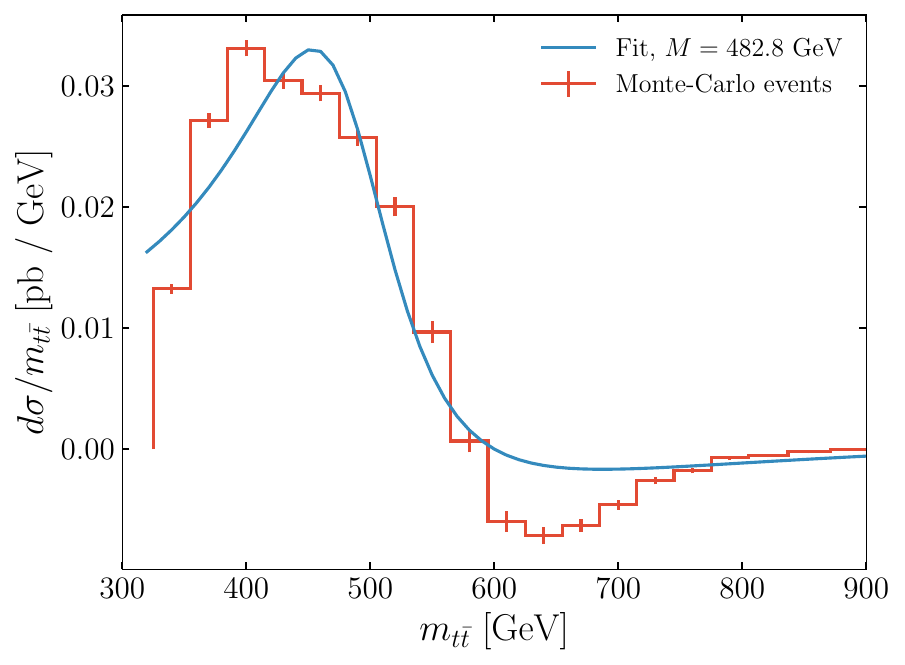}
    \caption{The individual contributions for the c2HDM point of~\cref{eq:benchmark_masscontributions} for $t \bar{t}$ production are shown on the left. On the right we show the sum of all the contributions and the curve for a peak--dip lineshape fit that is applied to the di-top final state.}
    \label{fig:fit}
\end{figure}

We next explore the bias that emerges for the case where a signal in the di-top final state originating from two near-degenerate scalars, which mix at loop level, is incorrectly interpreted. A scan is performed over different mass configurations, with $M_{h_1}$ denoting the physical (loop-corrected) mass of the lighter scalar. The couplings $c_{t_i}$ and $\tilde{c}_{t_i}$ are kept fixed to the values of~\cref{eq:benchmark_masscontributions}, while the loop-corrected masses, widths and the $\bm{Z}$-matrix are recalculated for each mass point. We subsequently generate the different resonant and interference contributions for each parameter point. The resulting signal with all the contributions is then fitted using a peak-dip lineshape parametrised as
\begin{equation}
    \label{eq:peakdipfit}
    \frac{1}{(m_{t\bar{t}} - M)^2 + (\Gamma/2)^2} \bigg[ A + B\,\bigg( (m_{t\bar{t}} - M) + C \,\frac{\Gamma}{2} \bigg)\bigg] - D \,,
\end{equation}
where $A, B, C, D, M$, and $\Gamma$ are fit parameters. This functional form provides an accurate description only in the vicinity of $m_{t\bar{t}} \approx M$. For a single scalar, $M$ would correspond to the physical mass of the resonance.  \Cref{fig:fit} illustrates this fit for the c2HDM parameter point of~\cref{eq:benchmark_masscontributions} featuring two scalars close in mass. The left panel shows the individual contributions to the invariant mass distribution; the right panel shows the fit to the sum of all contributions. The fitted mass is $\sim 100\gev$ below the average of the physical masses. Here, we implicitly assumed that the dip in the invariant mass spectrum is hard to resolve experimentally; therefore, the fit result is close to the peak position.

\begin{figure}
    \centering
    \includegraphics[height=0.7\linewidth,width=0.7\linewidth,keepaspectratio]{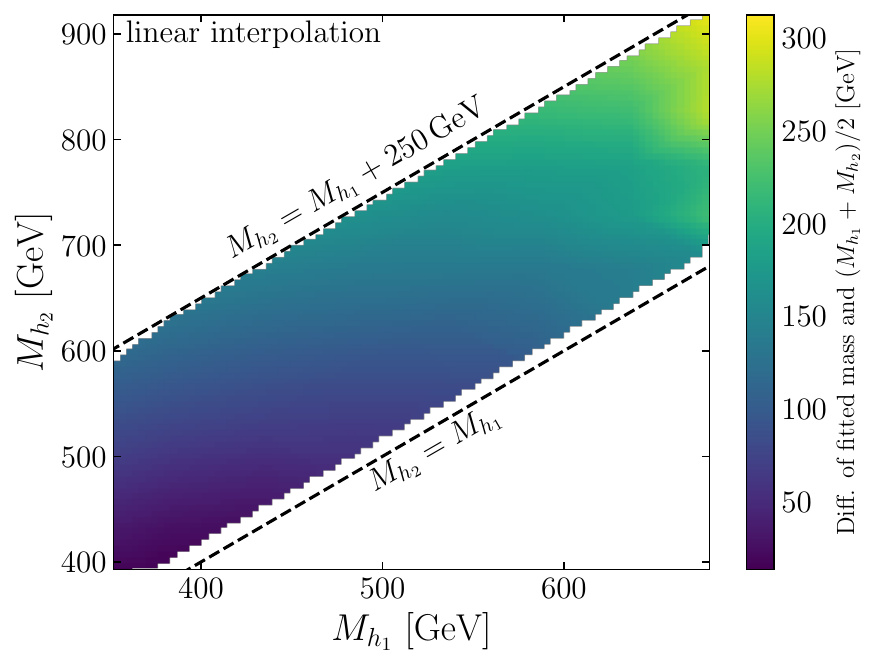}
    \caption{Difference between the mass value obtained from the peak--dip structure fit of the di-top final state and the average of the two scalar masses in the signal, $(M_{h_1} + M_{h_2})/2$. The difference is linearly interpolated between the simulated mass configurations.}
    \label{fig:2dmassdiff}
\end{figure}

For the performed scan we have restricted the scanned mass range of the heavier scalar to
\[
M_{h_2} \in (M_{h_1},\, M_{h_1} + 250~\text{GeV}) \,
\]
in order to focus on scenarios where the two masses are not vastly separated from each other. We have then carried out the fit for all parameter points and compared the fitted mass $M$ with the average of the two scalar masses. The resulting difference between the fitted mass $M$ and the average of the two scalar masses is displayed in~\cref{fig:2dmassdiff} in the plane of the two masses $M_{h_1}$ and $M_{h_2}$. While the shift is moderate for low $M_{h_1}$ and $M_{h_2}$, it increases to up to $300\gev$ when increasing both $M_{h_1}$ and $M_{h_2}$, as well as their mass difference. As explained above, the main origin is the interference with the background which shifts the signal peaks towards lower masses (a more sophisticated procedure for fitting a single peak--dip structure than the one used here could yield results that are closer to the actual physical masses). As a consequence, the observed mass difference can exceed the actual difference between the physical masses. These results illustrate the difficulty in associating a particular BSM signal with a deviation from the SM background that is detected in the di-top final state. While a broad peak starting at the $t \bar t$ threshold may be compatible with a variety of possible signals and thus would not be a reliable indicator for the physical masses of the particles causing the signal, the location of the dip is in fact directly correlated with the physical masses of the produced particles. However, the existence of a dip may be difficult to resolve in view of the experimental uncertainties of the analysis. On the other hand, if a peak arising from the four-top final state can be determined with sufficient statistical significance, it will not be affected by such ambiguities as a peak in the di-top final state and will therefore provide an important complementary probe allowing to pinpoint the underlying BSM model. 

\section{Conclusions}
\label{sec:conclusions}

Final states involving two or more top quarks are well-motivated search channels for BSM scalars. While the di-top final state has a larger cross-section than the four-top final state, a possible BSM signal in the di-top final state would be affected by interference contributions with the large QCD background of the di-top production process in the SM. This feature is present already for the production of a single BSM particle in the di-top final state. But for the case where two or more scalar BSM particles that can mix with each other contribute to the signal --- a situation occurring in many BSM scenarios --- even more complicated interference patterns can occur.

In this work, we have investigated the complementarity of the di-top and four-top final states for BSM searches, focussing for definiteness on the scenario comprising two BSM scalars with sizeable top-Yukawa couplings that can mix with each other. As a first step we showed explicitly, working with a toy model, that if both scalars are close in mass, the incorporation of loop-level mixing contributions is essential to get reliable predictions. If loop-level mixing is ignored, the cross-section is artificially enhanced, leading to a large overestimate of the experimental sensitivity and correspondingly of the obtained exclusion limits. We then confirmed that this behaviour also occurs for the di-top and four-top final states, employing realistic analysis setups.

Next, we investigated the interference patterns present in di-top and four-top production. Due to the much smaller SM background, signal--background interference has a significantly smaller influence in four-top production than in di-top production. Consequently, the invariant mass distribution for four-top production shows clear mass peaks that are associated with the physical masses of the produced BSM particles. The di-top invariant mass distribution, on the other hand, is affected by a large signal--background interference contribution, typically giving rise to a characteristic peak--dip signature. The peak starts at the $t \bar t$ threshold, can be rather broad and is difficult to relate to a specific underlying BSM scenario. While the location of the dip is correlated with the masses of the produced BSM particles, the existence of a dip may be difficult to resolve in the experimental analyses.

We have quantitatively compared the sensitivity of the di- and four-top final states. Due to the larger cross-section as well as the easier reconstruction, the di-top final state is significantly more sensitive than the four-top final state for detecting a deviation from the SM background. We showed, however, that the four-top final state can provide important complementary information. The di-top invariant mass distribution in the presence of a signal would often feature a broad peak above the $t \bar t$ threshold as a consequence of the large signal--background interference as well as the folding with the gluon parton-distribution-functions, which are largest close to the di-top threshold. Since the patterns of the broad peak would be compatible with a variety of possible BSM signals and the location of the dip may be difficult to resolve experimentally, the interpretation of a deviation from the SM background in the di-top final state would be highly non-trivial. In contrast, for four-top production a peak in the spectrum would be aligned with the physical masses of the produced particles. Therefore, a large excess in the di-top final state arising from a broad peak above the $t \bar t$ threshold accompanied by a smaller excess in the four-top final state from a peak at higher masses may point to the same underlying BSM scenario.

Our study highlights the importance of an accurate modelling of possible BSM effects and the importance of exploiting the complementary between different channels. We expect that future analyses combining di-top and four-top signatures, supported by improved theoretical predictions and advanced analysis techniques, will play a crucial role in uncovering or constraining extended scalar sectors at the LHC and future colliders.

\section*{Acknowledgements}

\sloppy{
H.B.\ is supported through the KISS consortium (05D2022) funded by the German Federal Ministry of Education and Research BMBF in the ErUM-Data action plan, by the Deutsche Forschungsgemeinschaft (DFG, German Research Foundation) under grant 396021762 --  TRR~257: \textsl{Particle Physics Phenomenology after the Higgs Discovery}, and through Germany's Excellence Strategy EXC~2181/1 -- 390900948 (the \textsl{Heidelberg STRUCTURES Excellence Cluster}). P.S. acknowledges support by the European Research Council (ERC) under the European Union’s Horizon 2020 research and innovation programme (Grant agreement No. 949451). K.P., P.G., R.K.\ and G.W.\ acknowledge support by the Deutsche Forschungsgemeinschaft (DFG, German Research Foundation) under Germany’s Excellence Strategy – EXC 2121 “Quantum Universe” – 390833306. This work has been partially funded by the Deutsche Forschungsgemeinschaft (DFG, German Research Foundation) - 491245950.}

\clearpage
\appendix

\section{Loop-level mixing --- additional results}
\label{app:eff_mix}

\begin{figure}[ht!]
    \centering
    \includegraphics[width=0.49\linewidth]{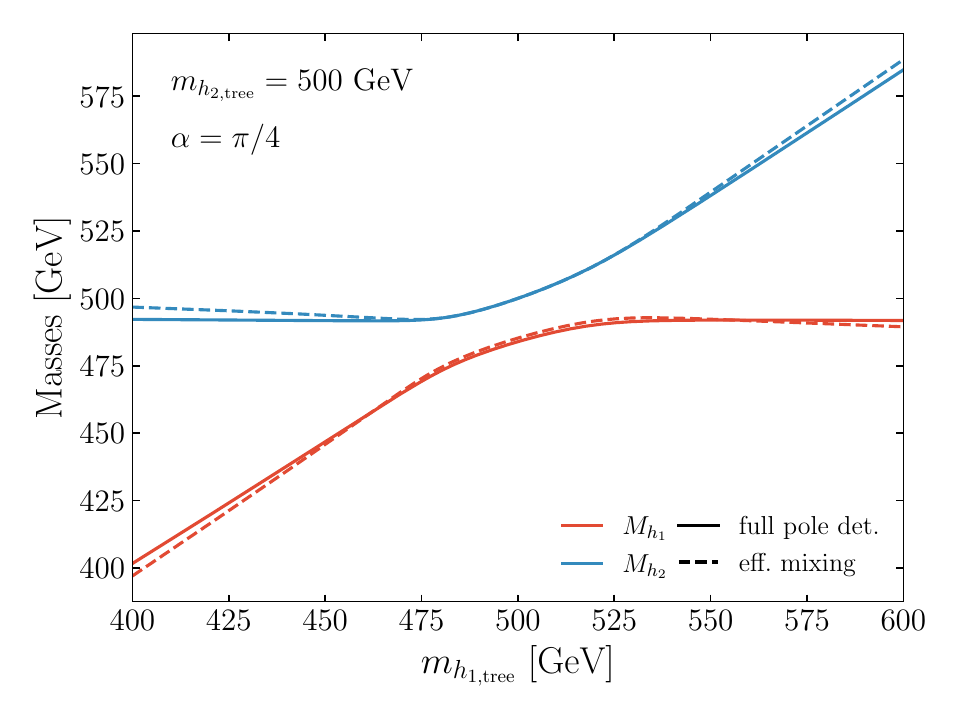}
    \includegraphics[width=0.49\linewidth]{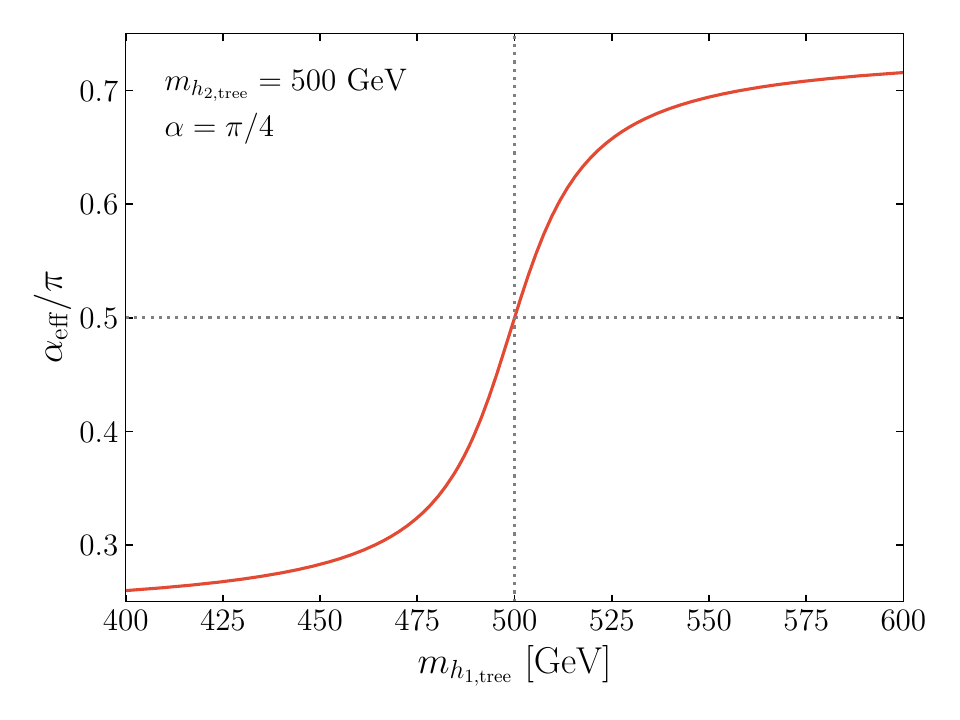}
    \caption{Left: Masses of $h_{1,2}$ as a function of the tree-level mass $m_{h_{1,\tree}}$ comparing the results of the full pole determination to the effective mixing angle approach. Right: Effective mixing angle as a function of $m_{h_{1,\tree}}$.}
    \label{fig:toy_masses_aeff}
\end{figure}

The loop-corrected masses of $h_{1,2}$ are shown in the left panel of~\cref{fig:toy_masses_aeff} as a function of $m_{h_{1,\tree}}$. We find that the effective-mixing-angle formalism provides a quite well approximation of the full result. 

For the effective mixing angle shown in the right panel of~\cref{fig:toy_masses_aeff}, we see that it varies from $\sim \pi/4$ for low $m_{h_{1,\tree}}$ masses to $\sim 3\pi/4$ for high $m_{h_{1,\tree}}$ masses. As shown in~\cref{sec:effective_mixing}, the effective mixing angle is exactly equal to $\pi/2$ if the two tree-level masses are equal to each other (as indicated by the dotted gray lines).

\clearpage

\section{Significance scans --- additional results}
\label{app:sig_scans}

\begin{figure}[ht!]
    \centering
    \includegraphics[width=0.47\textwidth]{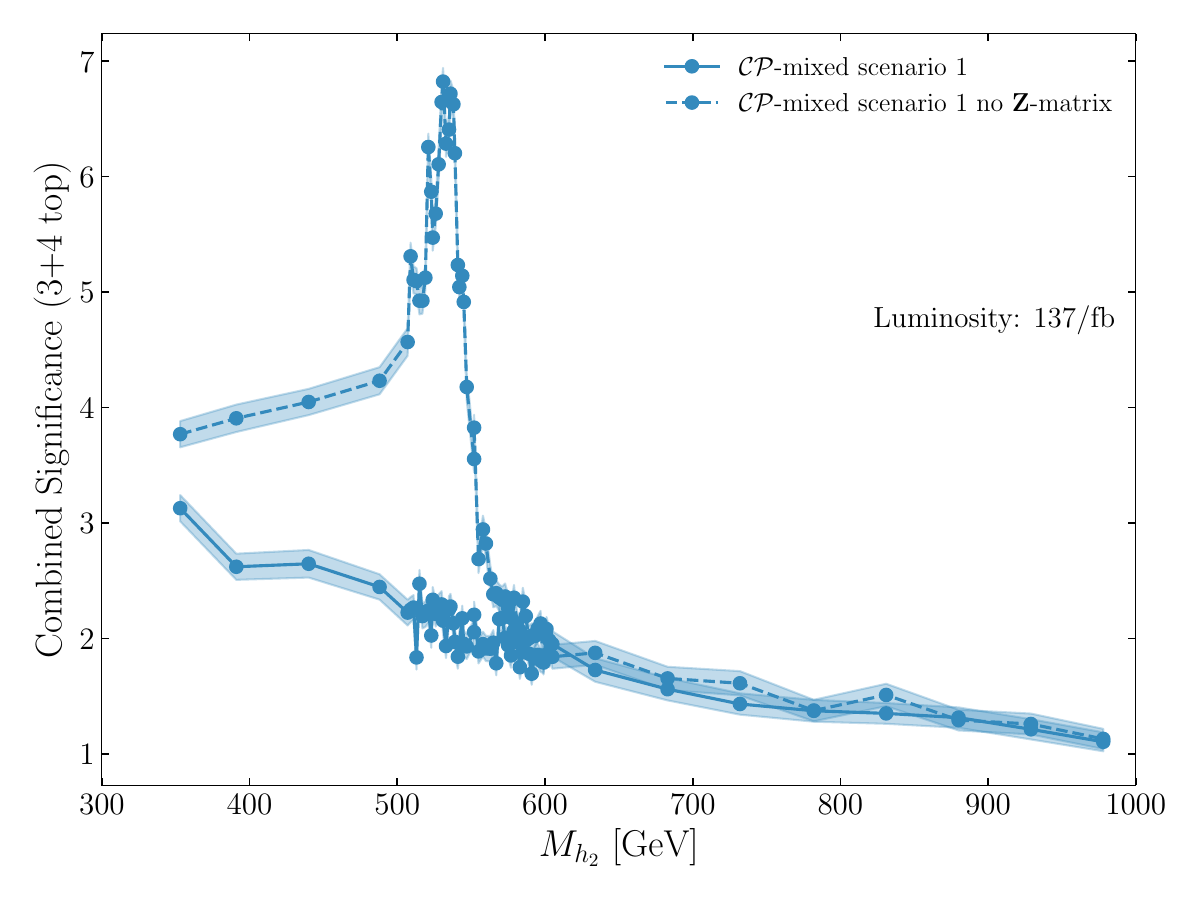}
    \caption{Three- and four-top significance as a function of the mass of the second scalar, $M_{h_2}$, comparing the \cp-mixed scenario~1 with and without incorporating the $\bm{Z}$-matrix. The shaded bands correspond to the uncertainty on the significance using 1000 pseudo-experiments in which signal and background have been sampled from normal distributions.}
    \label{fig:c2hdm-bp3-multitop-1a}
\end{figure}

We show in~\cref{fig:c2hdm-bp3-multitop-1a} the three-top and four-top discovery significance as a function of $M_{h_2}$ with the other parameters fixed according to the \cp-mixed scenario~1. We clearly see that ignoring the $\bm{Z}$-matrix leads to a large overestimate of the significance if $M_{h_1}\sim M_{h_2}$. If the $\bm{Z}$-matrix is correctly included, the significance curve shows a smooth behaviour up to numerical fluctuations.

\clearpage

\addcontentsline{toc}{section}{References}
\bibliography{biblio}

\end{document}